\documentclass[useAMS,usenatbib]{mn2e}
\usepackage{booktabs,graphicx,amssymb,epsfig,amsmath}
\usepackage{booktabs}
\newcommand{\changed}[1]{{\bf #1}}

\title[The inhomogenous reionization of the IGM by MPGCs]{The inhomogenous reionization of the inter-galactic medium by metal-poor globular clusters}
\author[B. F. Griffen et al.]{B. F. Griffen$^{1}$\thanks{E-mail: brendan.f.griffen@gmail.com} 
, M. J. Drinkwater$^{1}$, Ilian T. Iliev$^{2}$,  P. A. Thomas$^{2}$, Garrelt Mellema$^{3}$
\\ 
$^{1}$School of Mathematics and Physics, University of Queensland, QLD 4072, Australia\\
$^{2}$Astronomy Centre, University of Sussex, Falmer, Brighton BN1 9QH, UK\\
$^{3}$Department of Astronomy $\&$ Oskar Klein Centre, AlbaNova, Stockholm University, SE-10691 Stockholm, Sweden}

\begin{document}

\date{27 September 2012} 

\pagerange{\pageref{firstpage}--\pageref{lastpage}} \pubyear{2012}

\maketitle

\label{firstpage}

\begin{abstract}
We present detailed radiative transfer simulations of the reionization history of the Milky Way by metal-poor globular clusters. We identify potential metal-poor globular cluster candidates within the Aquarius simulation using dark matter halo velocity dispersions. We calculate the local ionization fields via a photon-conserving, three dimensional non-equilibrium chemistry code and allow the model to propagate through to the present day. The key feature of the model is that globular cluster formation is suppressed if the local gas is ionized.

We find that our spatial treatment of the ionization field leads to drastically different numbers and spatial distributions when compared to models where globular cluster formation is simply truncated at a given redshift. We investigate a range of plausible ionization efficiencies to determine the effect photon-rich and photon-poor models have on present day globular cluster properties. We find that it is possible for metal-poor globular clusters to have formed via the dark matter halo formation channel as our secondary model (delayed formation) combined with truncation at z = 10 produces radial distributions statistically consistent with that of the Milky Way metal-poor globular clusters.


If globular clusters do indeed form within high-redshift dark matter halos, if only in-part, their contributions to the reionization of the local (i.e. 2$^3$ $h^{-3}$ Mpc$^3$ centred on the host galaxy)  volume and mass by redshift 10 could be as high as 98$\%$ and 90$\%$, respectively. In our photon poorest model, this contribution drops to 60$\%$ and 50$\%$. This means globular clusters are important contributors to the reionization process on local scales at high-redshift until more photon-rich sources dominate the photon budget at later times. The surviving clusters in all models have a narrow average age range (mean = 13.34 Gyr, $\sigma$ = 0.04 Gyr) consistent with current ages estimates of the Milky Way metal-poor globular clusters.

We also test a simple dynamical destruction model and estimate that $\sim60\%$ of all metal-poor globular clusters formed at high redshift have since been destroyed via tidal interactions with the host galaxy. A final extension to our model utilises the Aquarius merger trees whereby  suppressed globular cluster descendants can potentially become active sources provided they reside within a neutral region of the simulation volume and have no active ancestors. This addition at least doubles the number of potential candidates but results in extended spatial distributions. 


\end{abstract}

\begin{keywords}
globular cluster -- radiative transfer: methods -- galaxy: formation -- cosmology: theory -- intergalactic medium
\end{keywords}

\section{Introduction}
It is now well understood that the universe emerged from the so-called \textit{dark ages}  (30 $<$ z $<$ 1100, \citealt{Rees97}) when light from the first stars and quasars ignited and radiated large quantities of ionizing photons into the inter-galactic medium (IGM). The two primary pieces of evidence we have of this process occurring are a) Ly$\alpha$ forest absorption spectra towards high-redshift quasars which show an increased opacity by the IGM at z = 6 (\citealt{Fan06}) and b) large-angle cosmic microwave background (CMB) polarisation anisotropies indicating that the median redshift of reionization was approximately z $\sim 10.4 \pm 1.4$ (\citealt{Komatsu09}). 

There are thought to have been two epochs of reionization: one for hydrogen (7 $<$ z $<$ 15, see \citealt{Loeb01} for a review), and one for helium (z $\sim$ 3.5, \citealt{Davidsen96}). Though the most likely sources of helium reionization are quasars, the sources of hydrogen reionization are far less well known. Possible sources of high redshift reionization include Population III stars (\citealt{Bromm04} and \citealt{Wise12} and references therein), Population II stars (e.g. \citealt{Sokasian03}), Quasars (\citealt{Madau99}, \citealt{Fan01}, \citealt{Fan06}) and more recently, particle decay/annihilation (\citealt{Mapelli06}). Recent work by \cite{Schaerer11} and \cite{Conroy11a} suggests that initial stellar masses of today's ancient globular clusters (hereafter GCs) could have been as much as 8-25 times higher when they first formed, reinforcing earlier conclusions that metal-poor GCs (hereafter MPGCs) could have significantly contributed to the reionization of the IGM at high redshift. \cite{Ricotti00} were among the first to formally suggest that GCs could have supplied a large quantity of the ionizing radiation but direct attempts to measure their contribution were carried out by \cite{Ricotti02}, \cite{Ricotti04} and \cite{Power09} (though \cite{Power09} were only examining X-ray binaries within GCs as means of reionizing the IGM).  Though insightful works, all of these studies unfortunately contain either (i) too many uncertain parameters (e.g. star formation efficiencies, escape fractions, photo-ionization rates) or (ii) lack the resolution required to accurately resolve globular cluster formation sites within their simulations. Despite these limitations, all of these studies have indicated that the flux from the first generation of globular clusters could have significantly contributed to the reionization of the Milky Way. 


The common approach used in many previous studies of GC formation (e.g. \citealt{Moore06}, \citealt{Bekki05}) was to essentially \textit{force} a truncation redshift (usually based on distant quasar observations) after which no more objects could form. This ensured that the correct satellite numbers and distributions were comparable to observations. Whilst this `reverse engineered' type technique is useful at describing what type of environment is required to align theory with observation, it does not account for the spatial inhomogeneity of the ionization field whereby sources are suppressed at different areas and at different times. \cite{Ciardi03}, \cite{Furlanetto04}, \cite{Barkana04} and \cite{Iliev06b} all clearly demonstrated that the reionization process of was in fact extended in time and spatially inhomogenous resulting in vastly different reionization times for different areas of the Universe. If studies of GC formation neglect the inhomogeneity of reionization field, their models will result in inaccurate formation numbers and distributions. 




Over the years, many have studied the connection between the reionization epoch and satellite formation. \cite{Alvarez09} for example, combined a $N$-body simulation with three-dimensional reionization calculations (1 $h^{-1}$ Gpc width) to determine the relationship between reionization history and local environment. They found that on average, halos with mass less than 10$^{13}$ M$_\odot$ were reionized internally, whilst almost all halos with mass greater than 10$^{14}$ M$_\odot$ were reionized from without (consistent with \citealt{Ocvirk11}). \cite{Busha10} combined the subhalo catalogs from the \textit{Via Lactea II} simulation with a Gpc-scale N-body simulation and found that by varying the reionization time over the range expected for Milky Way mass halos it could change the number of satellite galaxies by roughly 2--3 orders of magnitude. 

The work of \cite{Alvarez09} and \cite{Busha10} were semi-analytic studies of the reionization epoch. \cite{Iliev11} however, combined a cosmological simulation with radiative transfer and followed the formation of the Local Group of galaxies and nearby clusters. They found that for photon-poor models, the Local Group could have been reionized by itself (photon-rich models found that nearby clusters reionized the Local Group externally). \cite{Lunnan12} combined three-dimensional maps of reionization (using the semi-analytic models of \citealt{Furlanetto04}) with the initial density field of the \textit{Aquarius} simulation (\citealt{Springel08}). They found that the number of satellites depends sensitively on the reionization model, with a factor of 3-4 difference for a given host halo. Most recently and perhaps most intriguingly, \cite{Boylan-Kolchin11} have found that dissipationless dark matter simulations predict that the majority of the most massive subhaloes of the Milky Way are too dense to host any bright satellites at all. The vast majority of these works primarily focus on the effect of patchy reionization on the satellite (dwarf) systems of the Milky Way. All of them however, exclude the contributions from primordial GCs. Whilst all of these studies show the consequences of a patchy reionization process on galaxy evolution, a model incorporating both this process and \textit{all} of the relevant reionization sources is yet to be carried out.

\cite{Griffen10} (hereafter, G10) recently used cosmological simulations to study the reionization of the Milky Way by MPGCs using the Aquarius suite of simulations. They concluded that with a reasonable escape fraction and star formation rate, the primordial MPGCs of the Milky Way could have ionized the Milky Way by as early as  z $\sim$ 13. They found that the UV flux from high-z GCs had drastic consequences for not only the spatial and dynamical properties of present day (z = 0) GCs, but also for their overall ages as well. The main caveats of their work however, were i) they assumed that \textit{all} photons emitted from a potential MPGC would contribute to the reionization the Milky Way (much of the UV flux would have escaped into the outer IGM), ii) they did not model the dynamical destruction of the GCs once they merged with the central halo and iii) did not allow for delayed star formation to take place whereby a previously suppressed halo could reignite if the recombinations were high enough and/or it became neutral later.


In this paper we combine the same basic methods employed by G10 and addresses all of the major caveats. This is achieved by (i) modelling the patchy reionization process by calculating the propagation of ionization fronts \textit{directly} using a photon-conserving, three-dimensional ray tracing radiative transfer code ({\sc{C$^2$-Ray}}, \citealt{Mellema06}), (ii) measuring the effects of dynamical destruction on present day (z = 0) GC properties by including the Aquarius dataset with the dynamical models of \cite{Baumgardt03} and (iii) includes the effects of delayed star formation by combining the spatial information of halos within the Aquarius merger trees with the state of the IGM as modelled by {\sc{C$^2$-Ray}}. \textit{We aim to quantitatively analyse the formation of primordial GCs in a Milky Way type environment and determine to what extent GCs contributed to the reionization of the IGM and how the their contributions affected present day properties of metal-poor GCs}.


This paper is organised as follows: \S 2 describes the Aquarius dark-matter simulation, how we identify globular cluster candidates within Aquarius and the radiative transfer (RT) code used to model MPGCs as sources of reionizing radiation.  \S 3 describes the extensions to our model which include dynamical destruction and delayed star formation. \S 4 describes our results, exploring several scenarios to better understand the consequences of high and low escape fractions, halo concentrations, dynamical destruction and delayed star formation on present day (z = 0) properties of our GC candidate clusters. \S 5  discusses the consequences of this work in the wider context of galaxy evolution and large scale reionization. \S 6 describes the primary conclusions of this paper including a discussion of the caveats of this work and avenues for future study.

\section{Methodology}

In this section, we describe the details of our methodology. We begin by summarizing the relevant details of our dataset; the Aquarius simulation. We then describe how we identify globular cluster candidates using the Aquarius merger trees and lastly, we describe how we model GCs as ionizing sources. 

\subsection{The Aquarius simulations}

The Aquarius suite formally consists of six different simulations of Milky Way sized galaxies, one of which is the central dataset used in this paper. A more detailed description of the simulation can be found in S08, but we review the details relevant to this work here.

The Aquarius suite has same cosmological initial conditions as that of the Millennium simulation (\citealt{Springel05}). The initial size of the periodic box was 100 Mpc $h^{-1}$ Mpc set in a cosmology of  $\Omega_m$ = 0.25, $\Omega_\Lambda$ = 0.75, $\sigma_8$ = 0.9, $n_\mathrm{s}$ = 1, and Hubble constant $H_0$ = 100 $h$ km\ s$^{-1}$\ Mpc$^{-1}$ = 73 km\ s$^{-1}$\ Mpc$^{-1}$. We are aware that the value of $\sigma_8$ = 0.9 is higher than the currently accepted valued of $\sigma_8$ and could lead to an overestimate of the ages of all candidates since \textit{all} halos will form at earlier times. In terms of reionization, this means that for a fixed ionizing emissivity proportional to the collapsed fraction, the evolution is shifted to somewhat earlier times, resulting in an earlier overlap epoch (\citealt{Alvarez06}). Studies by \cite{Boylan-Kolchin11} predict that this would have a relatively minor effect.

The Millennium simulation was searched for halos of roughly Milky Way mass and without massive close neighbours in the present day (z = 0). \cite{Springel08} also checked that the semi-analytic modelling applied to the target haloes predicated them to host late-type galaxies. Otherwise the selection was random. These were then resimulated using $900^3$ particles in a box of dimension 100 $h^{-1}$ Mpc. After identifying the Lagrangian region from when each halo formed, the mass distribution was rerun at a much higher spatial and mass resolution. Although coarse particles were used to sample the distant regions, the resolution was such that the tidal field was accurately resolved at all times. 

The halos labelled, `Aq-A' to `Aq-F', stored snapshots at 128 output times equally spaced in log(a), where a = 1/(1 + z), represents the expansion factor (between redshift between z = 127 and z = 0). In this paper we only focus on the highest resolution halo available, AqA2. Dark matter haloes were identified using a combination of the friends-of-friends algorithm and {\sc subfind} (\citealt{Springel01}). As described in \cite{Cole08}, haloes and their substructure are traced through the snapshots and linked together in a merger tree. The smallest halo able to be resolved in the simulation is of order 10$^5$\ M$_{\odot}$. The most pertinent details of the Aquarius simulation used in this study are tabulated in Table \ref{tab:sim}. For greater detail on the simulation, see S08.

\begin{center}
\begin{table*}
\caption{The basic parameters of the Aquarius simulation data used in this paper.
}
\label{tab:sim}
\begin{minipage}{165mm}
 \begin{tabular}{cccccccccc}
\hline
Name & \textbf{$m_\mathrm{p}$} & $\epsilon$ & $N_{hr}$ & $N_{lr}$ &
$M_{200}$ & $r_{200}$ & $M_{50}$ & $r_{50}$ & $N_{50}$\\
& $[M_\odot]$ &[pc]& & & $[M_\odot]$ & [kpc] & $[M_\odot]$ & [kpc] & \\
\hline
Aq-A2 & $1.370 \times 10^4$ & 65.8 & 531,570,000 & 75,296,170 & $1.842 \times 10^{12}$ & 245.88 & $2.524 \times 10^{12}$ & 433.52 & 184,243,536 \\
\hline
\end{tabular}

{Notes: $m_p$ is the particle mass, $\epsilon$ is the
  Plummer equivalent gravitational softening length, $N_{hr}$ is the number of
  high resolution particles, and $N_{lr}$ the number of low resolution particles
  filling the rest of the volume. $M_{200}$ is the virial mass of the halo,
  defined as the mass enclosed in a sphere with mean density 200 times the
  critical value. $r_{200}$ gives the corresponding virial radius. We also give
  the mass and radius for a sphere of overdensity 50 times the critical density,
  denoted as $M_{50}$ and $r_{50}$. Note that this radius encloses a mean
  density 200 times the background density. Finally, $N_{50}$ gives the number
  of simulation particles within $r_{50}$.}
\end{minipage}

\end{table*}
\end{center}

\subsection{Identifying Primordial Globular Clusters}
\label{GCformmodel}
Independent of whether primordial GCs are significant contributors to the reionization of the IGM, the precise mechanism by which they form is still largely unknown. \cite{Peebles84} was one of the first to suggest that GCs may have formed within extended dark matter halos at high redshift. Observations of thin tidal tails by \cite{Grillmair95} and \cite{Odenkirchen03} drastically reduced the popularity of this model because it was found by \cite{Moore96}, that tidal tails should not exist if GCs live within extended halos. These studies however, assume that GCs at the \textit{present} period represent the environment from which they were born. There is currently no solid evidence to suggest that because GCs today aren't observed to contain significant traces of dark matter, that their more massive progenitors did not form within dark matter either.

 More recent work has renewed interest in the dark halo-GC formation scenario due to a range of numerical works published over the past decade (\citealt{Bromm02}, \citealt{Maschenko05}, \citealt{Bekki06}, \citealt{Bekki07}, \citealt{Boley09},  \citealt{Carretta10}). From a dynamical evolution standpoint, if GCs were in fact 8-25 times larger when formed than observed at present (\citealt{Schaerer11}), the likelihood of a giant molecular cloud ($f_{\mathrm{sfe}}\sim0.3$) of $\sim$ 10$^9$ M$_\odot$ collapsing \textit{without} being embedded within any dark halo is at this stage, small. \cite{Bromm02} and \cite{Maschenko05} have also shown that it is possible for the extended halos of GCs to be tidally stripped away by the present day. More recently, \cite{Boley09} carried out studies of the radial distribution of bright GCs in the Milky Way and concluded that it is possible that they formed in biased dark matter halos at high redshift. The formation of multiple stellar populations within GCs (\citealt{Bekki06}, \citealt{Bekki07} and \citealt{Carretta10}) are also consistent with the dark halo-GC formation channel.

In our work, we assume that high-z GCs \textit{did} form within the very first collapsed minihalos and due to dynamical disruption and violent relaxation, have since lost their dark matter halos. This assumption also makes it possible to model them as sources of reionization in current dark-matter only, $\Lambda$CDM simulations relatively straightforwardly. 

Our primary model of globular cluster formation is the same as that found in Section 2.2 of G10. In this work however, we focus on metal-poor GC production only but significantly improve on the previous work. We focus on metal-poor GC formation because their relative spread of ages (\citealt{DeAngeli05}) places them coincidentally at the ideal time to be significant contributors to the reionization process.

We adopt a relatively simple model to identify where the metal-poor GC objects would form within the Aquarius simulation based on the requirements for the collapse of a proto-GC gas cloud. The majority of previous studies adopt a similar model for GC formation, but the resolution of the Aquarius simulation allows us to \textit{directly} measure the primary parameter; temperature. 

\cite{Nishi02} and several others found that in order for gas clouds to efficiently cool, they must facilitate one or a combination of the three mechanisms; collisional excitation of hydrogen and helium, radiative recombination of hydrogen, and bremsstrahlung. The typical cooling function for a proto-GC candidate reveals an immense increase in the cooling rate as the temperature reaches 10$^4$ K, critical to the formation of stars.

Assuming the gas is in quasi-static equilibrium with the dark matter, we can use the virial theorem to relate the 1D velocity  dispersion of the dark matter halos to temperature. The ratio of the 1-D internal velocity dispersion of the dark matter subhaloes ($\sigma_v$) to the \textit{inferred} virial temperature of the gas, $T_v$ is given by;
\begin{equation}
\sigma_v^2=\frac{kT_v}{\mu m_{\rm H}},
\label{virialtemp}
\end{equation}
where \changed{$m_\mathrm{H}$} is the mass of a hydrogen atom and
$\mu$ is the mean molecular weight of the gas. We adopt molecular
weight of $\mu = 0.58$, appropriate for a fully-ionized, primordial
gas. Using these values, we therefore assume that whenever a halo's velocity dispersion increased \textit{above} 11.9 kms$^{-1}$ (T$_v$ $\sim$ 10$^4$ K), it is a \textit{GC candidate}. The smallest halo identified via this method contains 1619 particles, corresponding to a dark matter mass of 2.17 $\times$ 10$^7$ M$_\odot$. Throughout this work, `MPGCs' and `GCs' represent the exact same objects, metal-poor globular clusters.

\subsection{Treatment of GCs as ionizing Sources}
The Aquarius simulation allows us to identify where and when candidate sources (i.e. GCs) will form and our photon-conserving, three dimensional, non-equilibrium chemistry code ({\sc C$^2$-Ray}, \citealt{Mellema06}) will ensure the radiation from these sources is treated accurately.

\subsubsection{Radiative Transfer Calculations}
{\sc C$^2$-Ray}  is a grid-based short characteristics ray-tracing code which is photon conserving and traces rays away from our GC sources up to each cell. Photon conservation is assured by adopting a finite-volume approach when calculating the photoionization rates and by using time-averaged optical depths. Although helium is not modelled directly in this work, {\sc{C$^2$-Ray}} assumes that where and when hydrogen is ionized, helium is once (but not twice) ionized because of the similar ionization potentials of hydrogen and helium-I. The evolution of the ionized fraction of hydrogen, $x_\mathrm{i}$,  (including recombinations) is governed by,
\begin{equation}
\frac{dx_\mathrm{i}}{dt} = (1-x_\mathrm{i})(\Gamma_\mathrm{i} + n_\mathrm{e} C_\mathrm{i}) - x_\mathrm{i} n_\mathrm{e} \alpha_\mathrm{B}
 \end{equation}
where $n_e$ is the electron density, $\Gamma_{\mathrm{i}}$ is the photo-ionization rate, $C_{\mathrm{i}}$ is the collisional ionization rate and $\alpha_{\mathrm{B}}$ is the recombination rate. The code has been tested against 9 other rival ray tracing codes and was found to be on par with the best codes available today (\citealt{Iliev06b}, \citealt{Iliev09}). For more details, see \cite{Mellema06} and references therein.


 
A density mesh at both 256$^3$ and 512$^3$ resolutions of the AqA2 halo was created for each time step using a Cloud-in-Cell algorithm centred on the central host maintaining a physical volume of 6 $h^{-1}$ Mpc. This means that the cell width for our high resolution run is 11 $h^{-1}$ kpc/cell, significantly more accurate than the 195 $h^{-1}$kpc/cell used in \cite{Lunnan12}. Since the Aquarius simulation consists of both high and low resolution particles in the same volume, composite positions of all types of particles were used to ensure the density was properly represented at all spatial scales. 



\subsubsection{Mapping Sources}
\label{subsec:mapping}
Potential GC formation sites were located via the prescription previously discussed (see Subsection 2.2) through SQL queries of the publicly availably Aquarius database\footnote{Aquarius database;  http://www.galaxy-catalogue.dur.ac.uk :8080/Aquarius/}. These were then mapped into a grid based mesh for radiative processing according to the run's resolution. If more than one halo was identified within the same cell, those halos were combined and a luminosity assigned following a prescription based on their combined mass. For the 256$^3$ simulation, we found that only eight GC candidates formed within the same cell as another. When the resolution was increased to 512$^3$, no two GCs were found to have formed within the same grid cell (i.e. every GC was unique within its own grid cell).

The completeness of the identified sources is not homogeneous across the {\sc{C$^2$-Ray}} box. Since the Aquarius volume was resimulated in a Lagrangian volume, the total volume in which sources can be identified is \textit{less} than the total volume of the box used by {\sc{C$^2$-Ray}}. This incompleteness near the edges is yet another reason to restrict analysis to only those sources which have merged with the central halo. Despite having minor source incompleteness just near the edge of the {\sc{C$^2$-Ray}} volume, the density field in these outer regions is still accurately represented since the particles which created the mesh span the entire volume (high-resolution and low-resolution). 



\subsubsection{Calculating The Total Escape Fraction, $f_\gamma$}
As highlighted by \cite{Ricotti02} and \cite{Schaerer11}, the primary problem with all GCs studies is that there is significant uncertainty in the emissivity of GCs in the early universe. Exactly how many photons are produced in GCs over time depends largely on three parameters; what fraction of the gas is converted into stars $f_{\mathrm{sfe}}$, how effective are the stars at producing ionizing photons $\bar{q}$, which reach the intergalactic medium and finally, what fraction of the photons manage to escape the collapsing giant molecular cloud $f_{\mathrm{esc}}$. We parameterize all of these with a single parameter, $f_\gamma$ ($f_\gamma$ = $\bar{q}\times f_{\mathrm{sfe}}\times f_{\mathrm{esc}}$) which is equivalent to the number of ionizing photons/atom entering the IGM (between time steps). Since the code is sufficiently fast and the number of ionizing sources is relatively small, several runs of varying $f_\gamma$ values were able to be completed. 

\cite{Baumgardt07} found that in order to form a bound cluster, the star formation efficiency had to be one third or more. If all of the gas in a halo does not end up in the proto-cluster, the SFE could be considerably lower. The range of plausible values of the SFE adopted were, 0.15 $<$ $f_{\mathrm{sfe}}$ $<$ 0.7. 

\cite{Yajima11} recently combined three dimensional radiative transfer calculations and a cosmological smooth particle hydrodynamic simulation to study the escape fraction, $f_{\rm esc}$, of ionizing photons of high redshift galaxies. They found that the escape fraction drastically increased towards decreasing mass with halos of mass $<$10$^9$ M$_\odot$ having an average $f_{\rm esc} \sim 0.4$. As described in \cite{Ricotti02}, dust extinction and large levels of radiation are absorbed by the molecular cloud could drastically reduce this value. Our adopted escape fractions, ranging from photon-poorest to photon-richest scenario, were 0.1 $<$ $f_{\mathrm{esc}}$ $<$ 0.7.


As in G10, we use the the Population~II efficiency curve of Figure 2 from \cite{Tumlinson04} (calculated using the {\sc{STARBURST99}} code of \citealt{Leitherer99}) to calculate the number of ionizing photons emitted per baryon. Averaging over a Salpeter IMF (see Appendix of \citealt{Griffen10}) gives $\bar{q}\approx10\,000$. This figure could be raised by moving towards a more top-heavy IMF but recent work by \cite{Schaerer11} indicates that the ionizing photon output is quite independent of the IMF slope. 

Since there are relatively few observational constraints on the escape fraction and star formation efficiency at high redshift, the total escape fraction, $f_{\gamma}$, is largely a free parameter with significant uncertainty. However, by spanning a wide range of photon efficiencies, the consequences of having photon rich or photon poor metal-poor globular clusters on the reionization of the IGM can be examined without having to know what each of these quantities are \textit{exactly}. The upper and lower limits of the total number of photons per baryons being produced by GC candidates in this study can be found in Table \ref{tbl:paramters}. The total range of $f_\gamma$ values used in this study, ranging from extremely photon-poor to extremely photon-rich, are $f_\gamma$ = 150, 250, 500, 700, 1000 and 5000 photons/baryon. Our primary assumption here is that the baryons follow the underlying dark matter and photons are produced uniformly over a given time step. Any unsuppressed sources will continue producing photons in the subsequent time step. Each simulation has a time step of $\sim$20 Myrs.
  
In addition to the ionization models already described, we include a `truncation model' whereby halo formation is suppressed beyond a fixed redshift ($z_{\mathrm{trunc}}$). The only other requirement for these halos is that they have merged with the central halo by the present day (z = 0).  G10 originally used this type of suppression model and found a $z_{\mathrm{trunc}}$ = 13 to be the most suitable of the possible truncation redshifts. This truncation redshift can ultimately be arbitrary as it simply depends on the efficiency of the sources used. We adopt $z_{\mathrm{trunc}}=13$ here for the simple purpose of illustrating the effect inhomogeneous ionization models have on the numbers and distributions of GC candidates when compared to a truncation models.

\subsubsection{Suppression Criterion}
Throughout this work we define `suppression' as the inability of candidate halos to develop into GCs because their gas content is ionized \textit{locally}. A radiative transfer cell is considered ionized when the ionized fraction of hydrogen is above a fixed threshold ($x_{\mathrm{thresh}}$). Throughout our entire study we set $x_{\mathrm{thresh}}$ = 0.1. GC candidates residing in a cell with an ionized fraction above this threshold are suppressed.

An increase in the ionized fraction threshold essentially decreases the strength of the suppression process. To gain a full understanding of the subtle differences in the strength of the suppression on the overall reionization process, a range of ionized fraction thresholds must be tested. In Appendix B of \cite{Iliev11b}, they discuss how source suppression depends on the Jeans mass suppression threshold. They found both $x_{\mathrm{thresh}}$ = 0.1 and 0.5 yield essentially the same IGM evolution, apart from a slight offset in time. Only for $x_{\mathrm{thresh}}$ = 0.9 (weak suppression) does the evolution of the IGM begin to behave very differently. However, they were studying the reionization process in large scale galaxy cluster environments and so these inferences do not necessarily translate directly to a smaller Milky Way type environment like the one studied here. The only way to determine the effect of higher ionized fraction thresholds is to do a direct recalculation of one of the runs presented here but with higher threshold values. We defer such a study to future work (see Section 5 for further discussion). In any case, in \textit{this} work, we only require that a halo is for the first time above $\sigma > 11.9$ km/s and that it resides within a cell with an ionized fraction, $x < 0.1$.

In terms of computation time, the highest resolution simulations (512$^3$, Model 2: delayed) were carried out in parallel on 8 nodes, each using 8 cores, requiring 1.2 GB of memory (for all sources) and 9 hours of computation time. The lowest resolution simulation (256$^3$, Model 1: permanently suppressed) took 1 hour of computation time.

\begin{table}
 \centering
 \small
   \caption{ The photon-poorest and photon-richest scenarios constrained by previous computational studies. These limits  set the upper and lower bounds on the total number of photons/baryon leaving a proto-GC, $f_\gamma$, in this study. }
 \begin{minipage}{70mm}
 \begin{center}
  \begin{tabular}{ccc}
\hline
\hline
Parameter & Photon-poorest  & Photon-richest   \\
 & Model & Model \\
\hline
\hline
$\bar{q}$ & 10,000 ph/b  &  10,000 ph/b\\
$f_{\mathrm{sfe}}$ & 0.15  &  0.7\\
$f_{\mathrm{esc}}$ & 0.1  &  0.7\\
\hline
$f_\gamma$ & 150 ph/b & $\sim$5000 ph/b\\
\hline
\hline
\end{tabular}
\end{center}
 \label{tbl:paramters}
\end{minipage}
\end{table}

\section{Model Extensions: Delayed Star Formation $\&$ Dynamical Destruction}

In this section, we detail two extensions of the previous model; delayed star formation and dynamical disruption. Whilst these processes have been studied before, it is the first time that the following two models have been combined together in the context of GC formation.

\subsection{Delayed Star Formation}
\label{subsection:delayedmodel}
As an extension to the previous model,  we made use of the merger trees to trace candidate GCs spatial and dynamical evolution throughout the AqA simulation. In many previous works, the suppression of candidate clusters is treated in a simple, straightforward manner. If a candidate satellite resides within an ionised region, it is suppressed permanently. We too adopt this technique for some of our models and class them as `Model 1' or the `permanently suppressed' model.

On cosmological scales, such `suppressed once-and-forever' type models are a good approximation since there are many neighbouring sources to keep most regions ionised. Any descendent of a suppressed halo entering one of these ionised regions will remain suppressed. On small scales however, such as the one being dealt with here, there are occasions where a candidate GC, previously suppressed, may be able to become active again once its gas content becomes neutral at a later time-step. This process effectively delays the star formation of a GC candidate and adds previously suppressed candidates to the list of legitimate GCs in the present day (z = 0). This important process can be closely examined by using the merger trees of the simulation under study. In this work, we denote this type of model as `Model 2' or the `delayed' model. These names are used interchangeably though out this work.

Our delayed model works as follows. At each snapshot, we add the descendent IDs of each of the active halos to a cumulative list. In the subsequent snapshot, candidate halos are checked against this list to determine if any of the potentially active sources have been active before. If a match is found, this halo's descendent ID is added to the list and it is removed from the potential new sources for that snapshot. If a match is not found, the ionisation state of the cell that source resides in is checked and if it is neutral (or below $x_{\mathrm{thresh}}=0.10$) it will then become active and the usual quantities propagated. 

In this way, Model 2 has two types of populations forming over the course of the simulation; `first timers' (i.e. source which becomes active in the same snapshot they are identified) and `delayed' (i.e. they become active at a later snapshot than the one they are identified). This also self-consistently excludes any halos which have \textit{any} active ancestors. At early times, one would expect there to be direct overlap of the total number of sources using either Model 1 and Model 2 but once delayed halos add substantial ionising photons to the IGM at later times, this will alter the number and present day properties of the candidates across the two models.

In summary, Model 1  (or M1) represents a `permentantly suppressed'  scenario whereby ionization suppresses a GC candidate permentantly and \textit{never} ignites at any subsequent time step. Model 2 (or `delayed model') incorporates all of the same identification methods used in Model 1 except that now the descendents of previously suppressed halos can become active if their gas content becomes sufficiently neutral and they have no active ancestors.

In this work we refer to Model 1 with $f_\gamma$ = 500 (where $f_{\mathrm{esc}}$ = 0.2 and $f_{\mathrm{sfe}}$ = 0.25) as our fiducial model because the adopted total escape fraction is within the reasonable range of literature estimates.  Table \ref{tbl:simulationlist} lists the two types of models tested and the range of photoionisation efficiencies adopted in this work (e.g. our fiducial model is represented by $\mathrm{M1\_512\_ph500}$; Model 1 at 512$^3$ resolution with $f_\gamma = 500$).



\begin{table}
 \centering
 \small
 \begin{minipage}{70mm}
  \caption{The simulation suites used in this study. `M1' and `M2' denote Model 1 (permanently suppressed model) and Model 2 (delayed model), respectively. Our fiducial model is $\mathrm{M1\_512\_ph500}$. A lower resolution run ($256^3$) of our fiducial model model is also carried out to test  for convergence to ensure our results are not resolution dependent (see Sub-section \ref{subsec:convergence}) .}
 \begin{center}
  \begin{tabular}{ccc}
\hline
\hline
 Name &  $\#$ cells &  $\mathrm{f_{\gamma}}$ [ph/b]  \\
\hline
\hline
M1\_512\_ph150 & 512$^3$  &  150\\

M1\_512\_ph250 & 512$^3$  &  250\\

M1\_256\_ph500 & 256$^3$  &  500\\
\textbf{M1\_512\_ph500} & \textbf{512$^3$}  &  \textbf{500}\\

M1\_512\_ph700 & 512$^3$  &  700\\

M1\_512\_ph1000 & 512$^3$  &  1000\\

M1\_512\_ph5000 & 512$^3$  &  5000\\

\hline
M2\_512\_ph150 & 512$^3$  &  150\\

M2\_512\_ph250 & 512$^3$  &  250\\

M2\_512\_ph500 & 512$^3$  &  500\\

M2\_512\_ph700 & 512$^3$  &  700\\

M2\_512\_ph1000 & 512$^3$  &  1000\\

M2\_512\_ph5000 & 512$^3$  &  5000\\
\hline
\hline
\end{tabular}
\end{center}
 \label{tbl:simulationlist}
\end{minipage}
\end{table}

\subsection{Dynamical Destruction}
\label{subsec:dyndest}
In the second extension to our model, we address the important process of dynamical destruction. Although we have excellent resolving power to locate where candidates form, it is impossible to locate these halo structures in the present day (z = 0) as the vast majority of them will have undergone mergers with either the host or each other. As with G10, we assume the collapsed baryonic gas fraction of a GC at the centre of each halo will follow the same trajectory as that of the most bound particle of the halo. At each snapshot we simply identify the most bound particle corresponding to each GC candidate. We then follow this most bound particle through to the present day allowing us to track the GC formation sites over the course of the entire simulation. 

During mergers with the central halo, the formation site is no longer able to be tracked by {\sc{subfind}} and so we must infer the trajectory of the formation site within the host halo based on the most bound particle alone.  \cite{Baumgardt03} derived a fitting formula based on a series of $N$-body models to determine whether a star cluster, given a few basic orbital characteristics, will be disrupted during its orbit around a host after such an encounter takes place. The expression they derived (Equation 10 from \citealt{Baumgardt03}) for the dissolution time was, 
\begin{equation}
T_{\mathrm{diss}} = \beta \left[\frac{M}{0.33\ M_\odot}\right]^x \frac{R_\mathrm{G}}{\mathrm{kpc}} \left(\frac{V_\mathrm{G}}{\mathrm{220\ km\ s^{-1}}}\right)^{-1} (1 - \epsilon), \\
\end{equation}
where $M$ is the halo mass before infall, $\epsilon$ is the orbital eccentricity, $R_\mathrm{G}$ is the apocentric distance of the halo (kpc), $V_\mathrm{G}$ is the circular velocity (km/s) of the host galaxy and $x$ and $\beta$ are cluster scaling law parameters depending on the cluster concentration.  From a fit to the lifetimes of the \cite{King66} models with dimensionless central potential W$_0$=5.0, \cite{Baumgardt03} found $x = 0.75$, $\beta = 1.91$. 


For more concentrated, $W_0 = 7.0$ King profile clusters, they found a small change in the dissolution time of a cluster (see Fig. 3 of \citealt{Baumgardt03} for details) and so for the sake of simplicity we restrict our dissolution analysis to W$_0$ = 5.0 King profile clusters only.  In order to determine whether a candidate GC survives an encounter with the central host halo, its calculated dissolution time must be compared to the time it merged with the central host. If the time since the candidate GC merged with the central halo exceeds its dissolution time, then the cluster is disrupted.



\section{Results}
\label{sec:results}
Here we present the results of our simulations organised as follows. Section 4.1 outlines the results for Model 1. Section 4.2 uses the same candidates identified in Section 1 but applies the dynamical disruption model outlined in Section 3.2. Finally, Section 4.3 presents the results of Model 2 which includes previously suppressed sources which are allowed to turn on given the correct environmental and antecedent conditions. The objects used in all of the following analyses have been restricted to only those which merge with the final host by z = 0. Specifically, \textit{merged} means that {\sc subfind} can no longer recognise them as two separate structures in the present day. The total number of objects formed without suppression is 310. Though there are objects which retain their dark matter halos, we cite G10 as a preliminary analysis of these objects and save more detailed study of these survivors for future work. 

\subsection{Model 1: `suppressed permentantly'}
\subsubsection{Visualisation Of The ionization Field}
\begin{figure*}
\centering
\begin{tabular}{cccc}
\includegraphics[width = 0.24\textwidth]{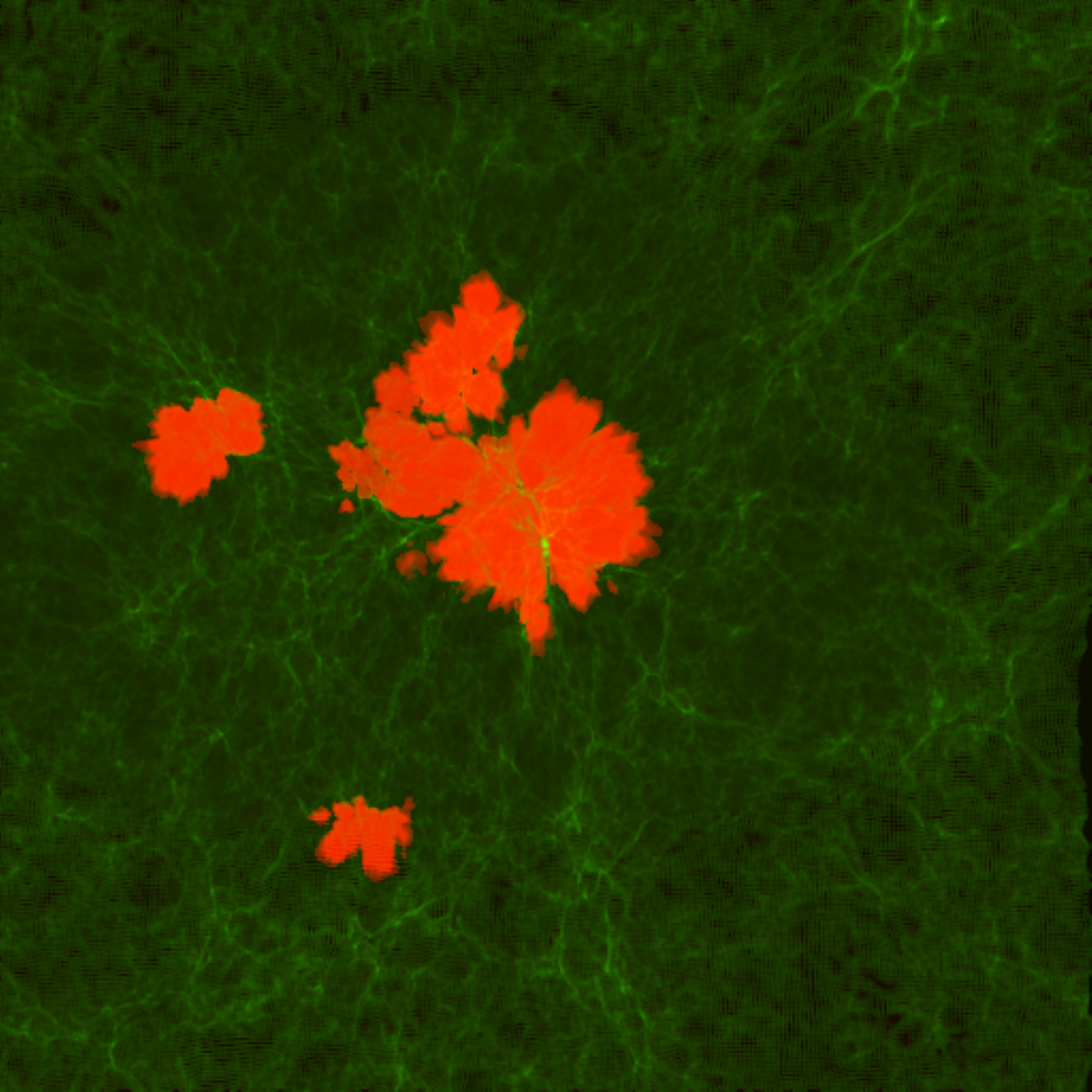} 
\includegraphics[width = 0.24\textwidth]{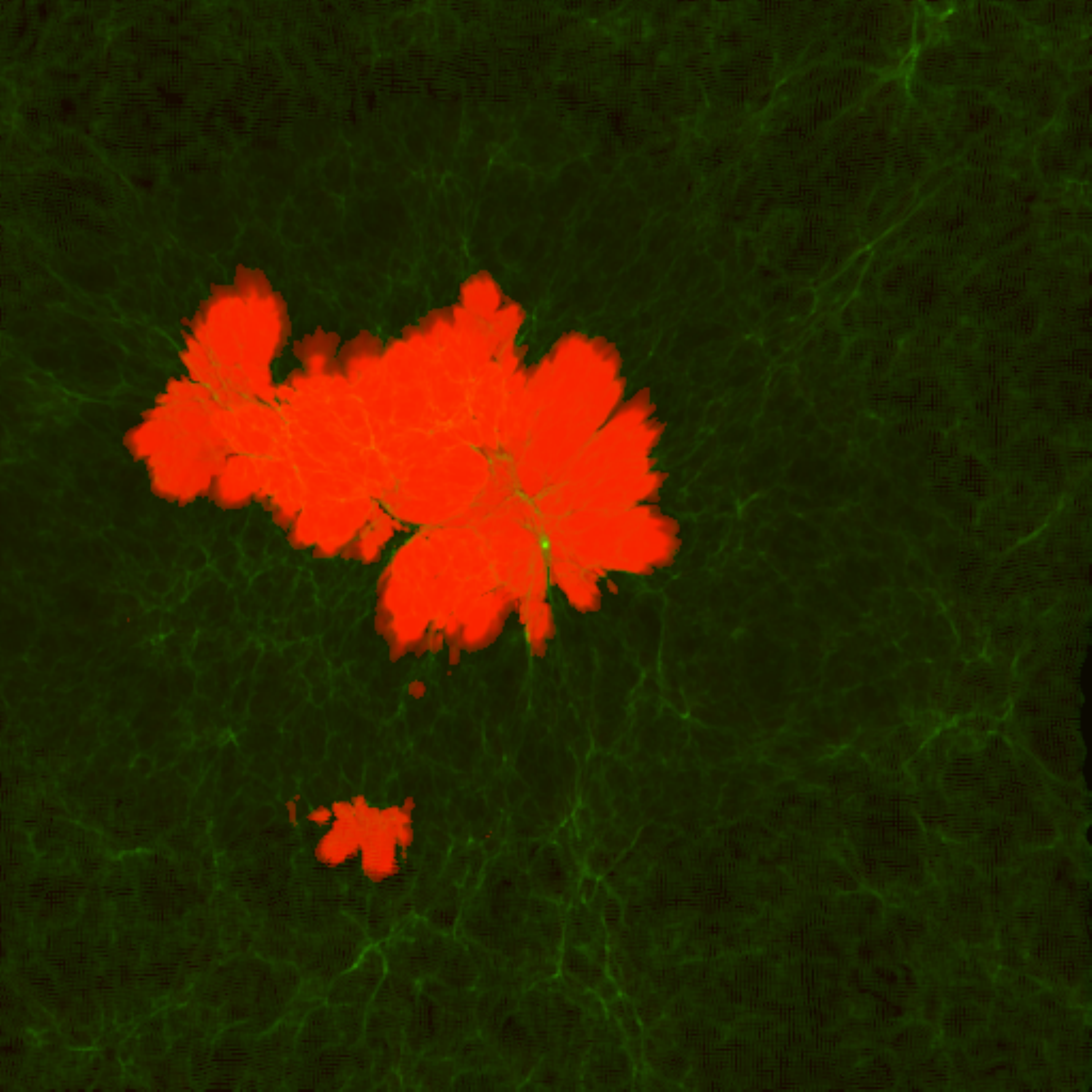} 
\includegraphics[width = 0.24\textwidth]{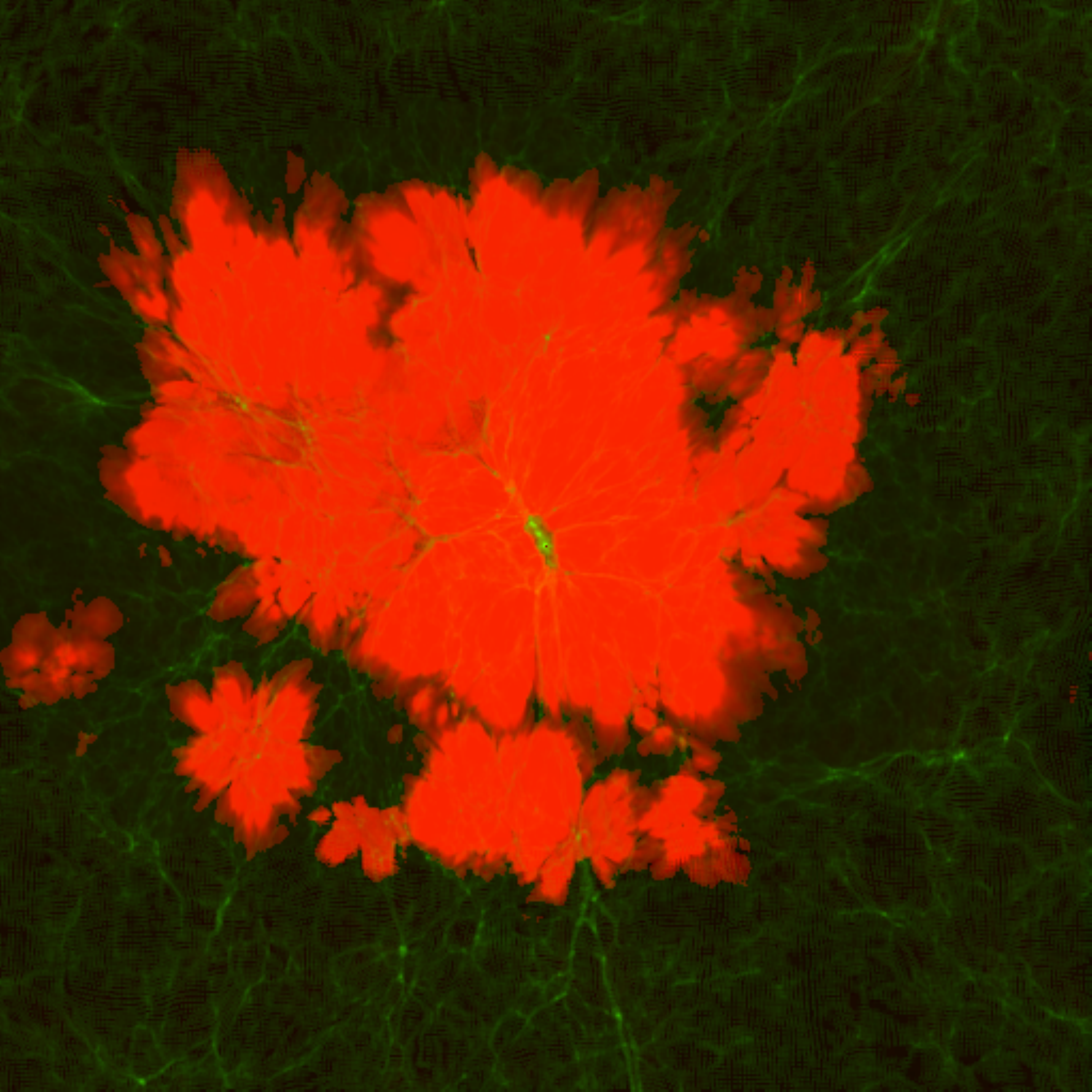}
\includegraphics[width = 0.24\textwidth]{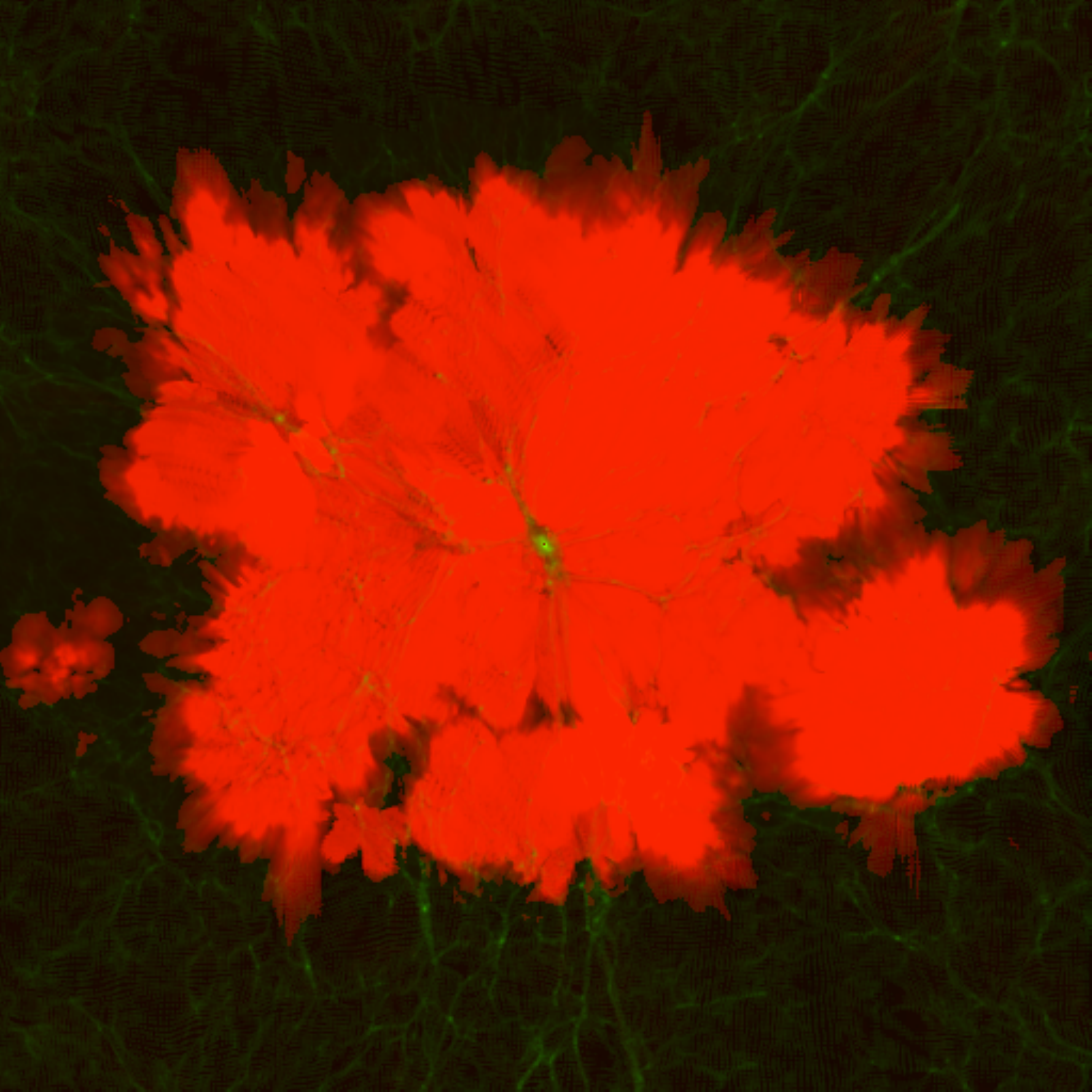}
\end{tabular}
\caption{Time evolution of the ionization field. Each panel represents a spatial slice (x-y projection), 460 $h^{-1}$kpc thick, of the ionized and neutral gas density from simulations of Model 1 using photo-ionization efficiencies of $f_{\gamma}$ = 500 at z = 14.03, 12.07, 8.90 and 7.03 respectively. The box in each panel has a comoving width of 6 $h^{-1}$ Mpc.}
\label{fig:same_fgamma}
\end{figure*}

\begin{figure*}
\centering
\begin{tabular}{cccc}
\includegraphics[width = 0.31\textwidth]{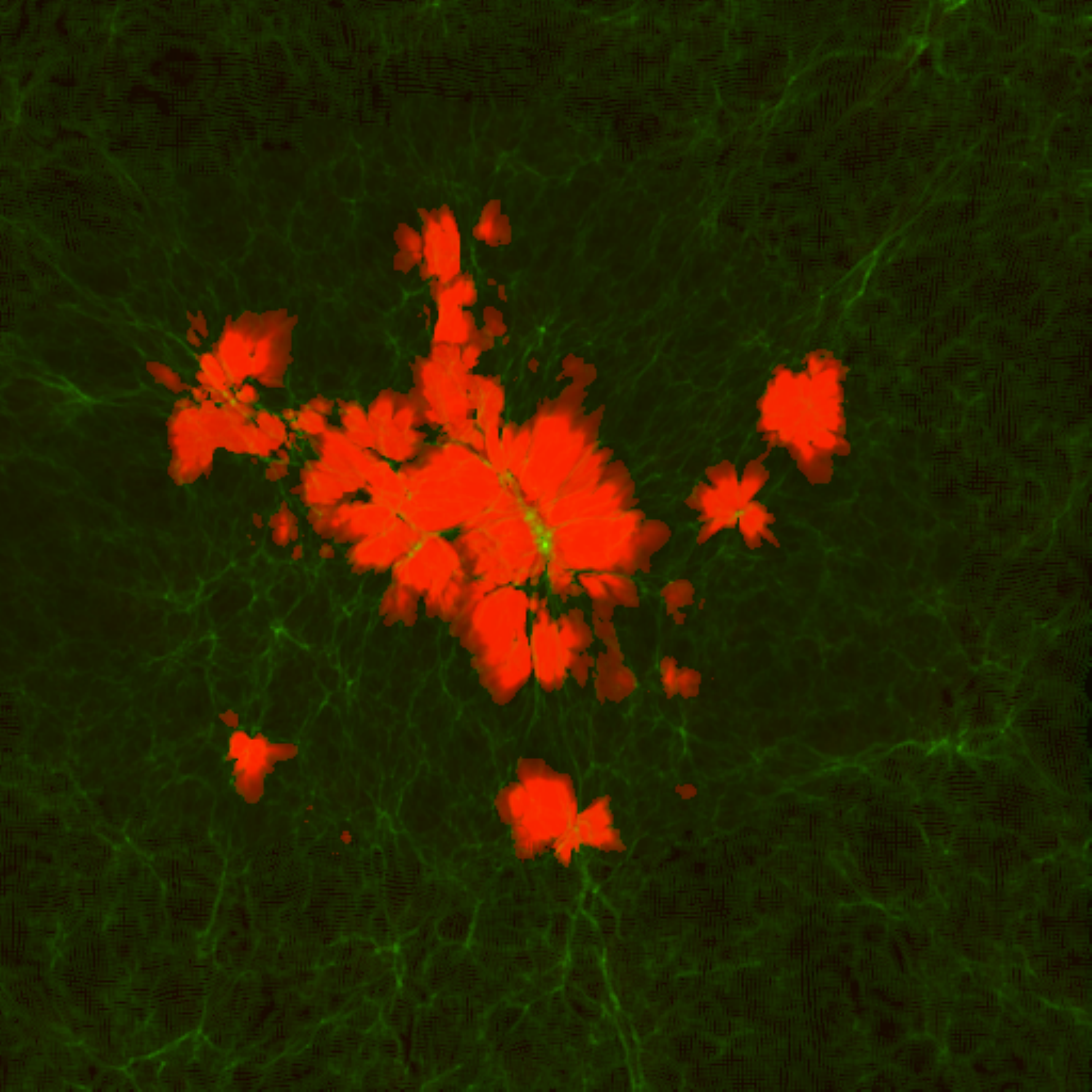} &
\includegraphics[width = 0.31\textwidth]{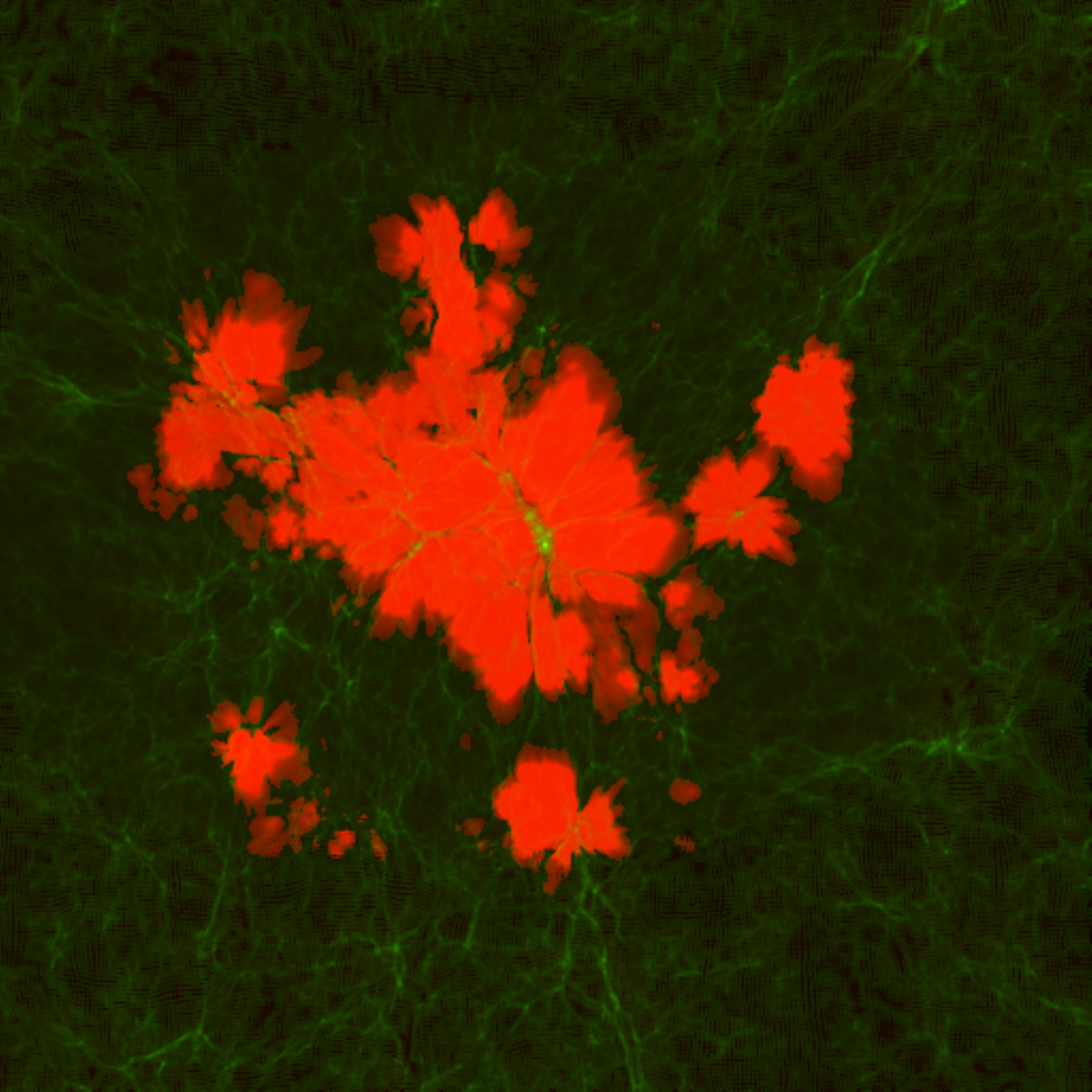} &
\includegraphics[width = 0.31\textwidth]{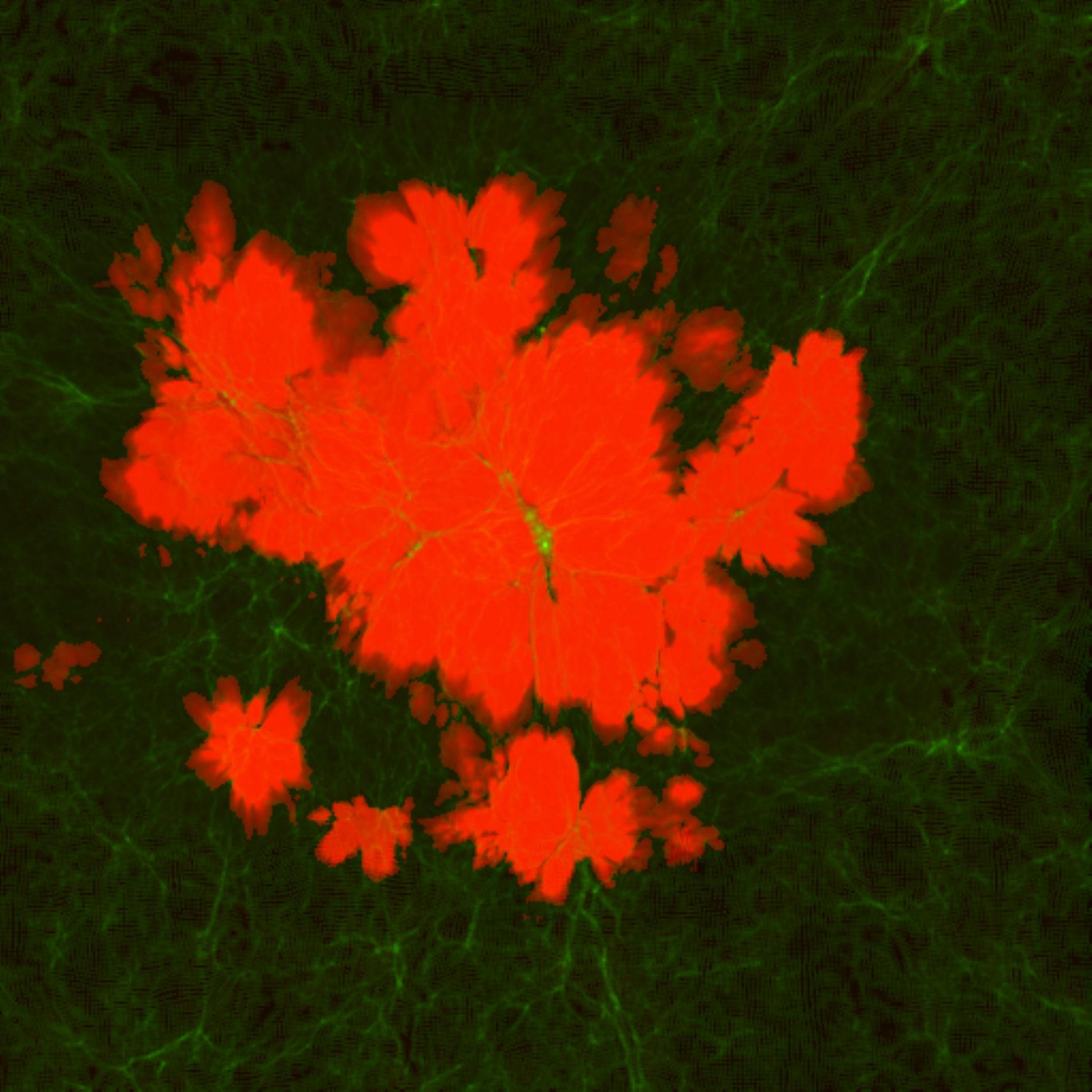} \\ \\
\includegraphics[width = 0.31\textwidth]{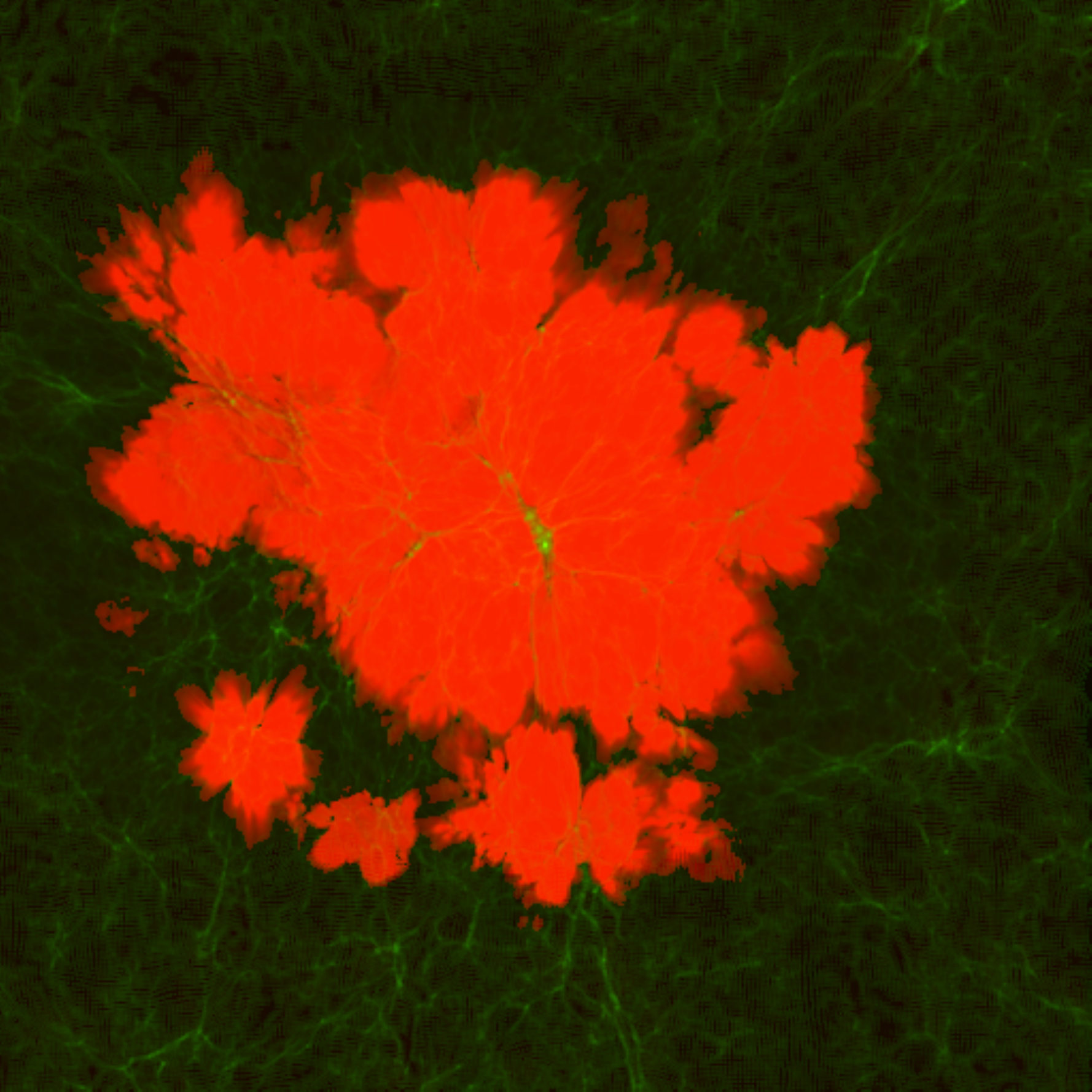} &
\includegraphics[width = 0.31\textwidth]{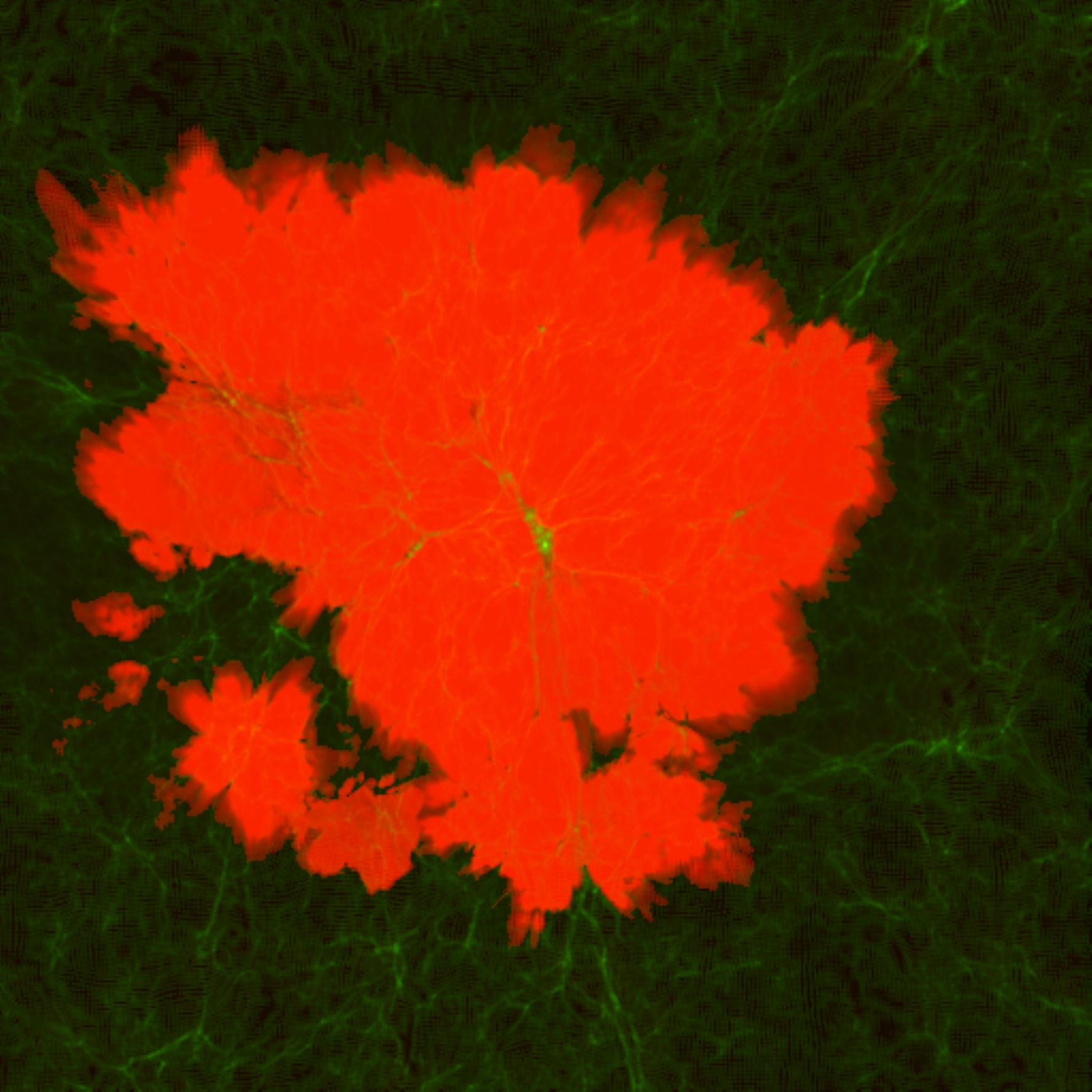} &
\includegraphics[width = 0.31\textwidth]{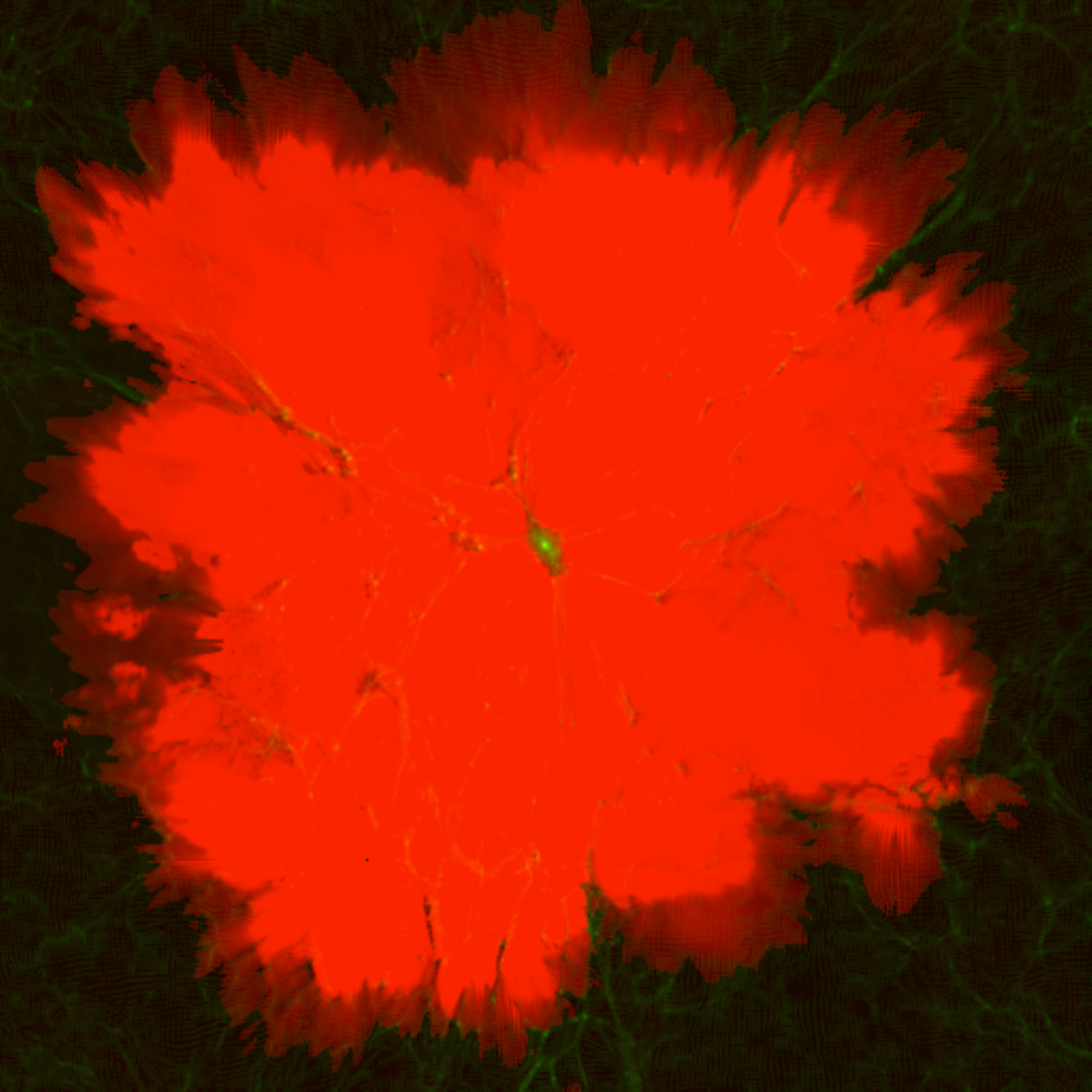} &
\end{tabular}
\caption{Snapshots of the ionization field for the six photo-ionization efficiencies tested in this study all taken at the same redshift (z = 10.11). Again, each panel represents a spatial slices (x-y projection), 460 $h^{-1}$kpc thick, of the ionized and neutral gas density from simulations of Model 1. Top left to top right: $f_\gamma$ = 150, 250, 500 photons/baryon and bottom left to bottom right:  $f_\gamma$ = 700, 1000 and 5000 photons/baryon. The box in each panel has a comoving width of 6 $h^{-1}$ Mpc.}
\label{fig:diff_fgamma}
\end{figure*}

We identify GCs using the method described in Section \ref{GCformmodel} and show a visual representation of these candidates ionizing the IGM in Figures \ref{fig:same_fgamma} and \ref{fig:diff_fgamma}. Figure \ref{fig:same_fgamma} shows the neutral gas density using a fixed emissivity of f$_\gamma$ = 500 photons/baryon at z = 14.03, 12.07, 8.90, 7.03 within a box of 6$^3$ $h^{-3}$ Mpc$^3$. This shows a clear evolution of the ionizing fronts throughout the IGM during the formation of the proto-Milky Way particularly highlighting the inhomogeneous nature of the ionization field. Figure \ref{fig:diff_fgamma} shows six runs at differing emissivities all captured at the same time (z = 10.11). The emissivities range from f$_\gamma$ = 150 to f$_\gamma$ = 5000 and clearly illustrates how higher photo-ionization efficiencies ionize larger volumes. Interestingly the very central regions of the host galaxy remain relatively neutral (R $<$ 100 kpc). An increase in the local recombination rate ($\propto \rho^2$) in this region makes it difficult for central region to become ionized. 


\subsubsection{Formation History $\&$ Numbers}
\begin{figure*}
\centering
\includegraphics[width = 1\textwidth]{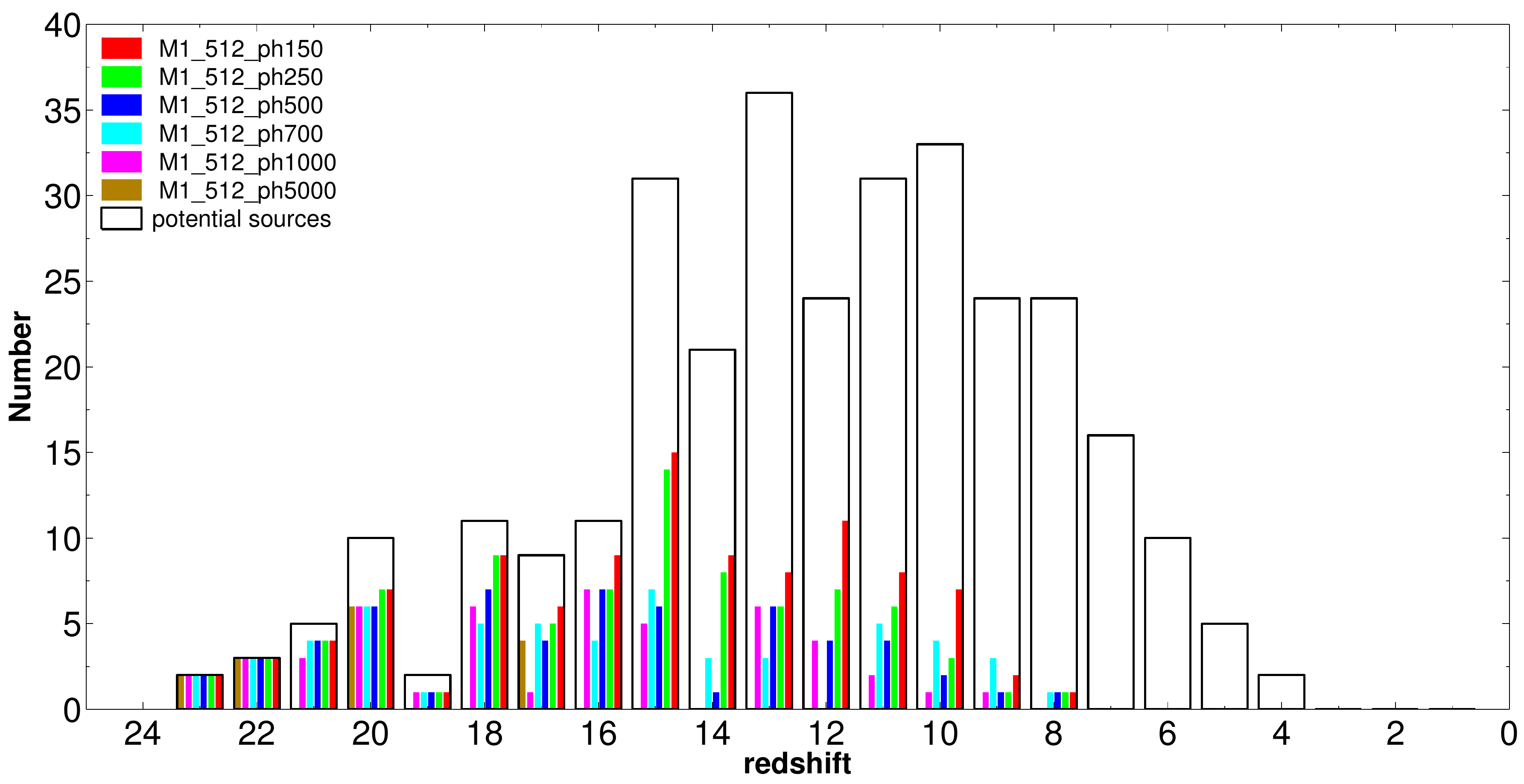}
\caption{Formation histograms for each of the respective ionization efficiencies. The number of objects as a function of redshift are binned in whole integer redshifts for each snapshot of the simulation. For ever higher ionization efficiencies, more photons escape into the IGM, resulting in fewer objects forming at lower redshift (i.e. only those identified by the velocity-temperature criterion from Aquarius) due to to an increased level of suppression. The clear histogram represents the formation of GCs without suppression included.}
\label{fig:formation_hist}
\end{figure*}

The formation history and overall numbers in each of our models are shown in Figure \ref{fig:formation_hist}. GC formation \textit{without} suppression is shown by the open bins outlining each of the suppression models (shown in colour).  As expected, the relative numbers of objects at high-redshift are similar as the formation environments are roughly the same. As the activation of more sources proceeds, large quantities of ionizing photons are injected into the IGM resulting in greater suppression in the higher escape fraction models than the lower escape fraction models. This amount of suppression taking place is most significant at z $<$ 16. There are no active sources in any of the models below  z = 8 except for one lone source forming at z = 5 in the photon-poor, M$1\_512 \_\mathrm{ph}150$ model. 

\begin{figure}
\centering
\includegraphics[width = 0.45\textwidth]{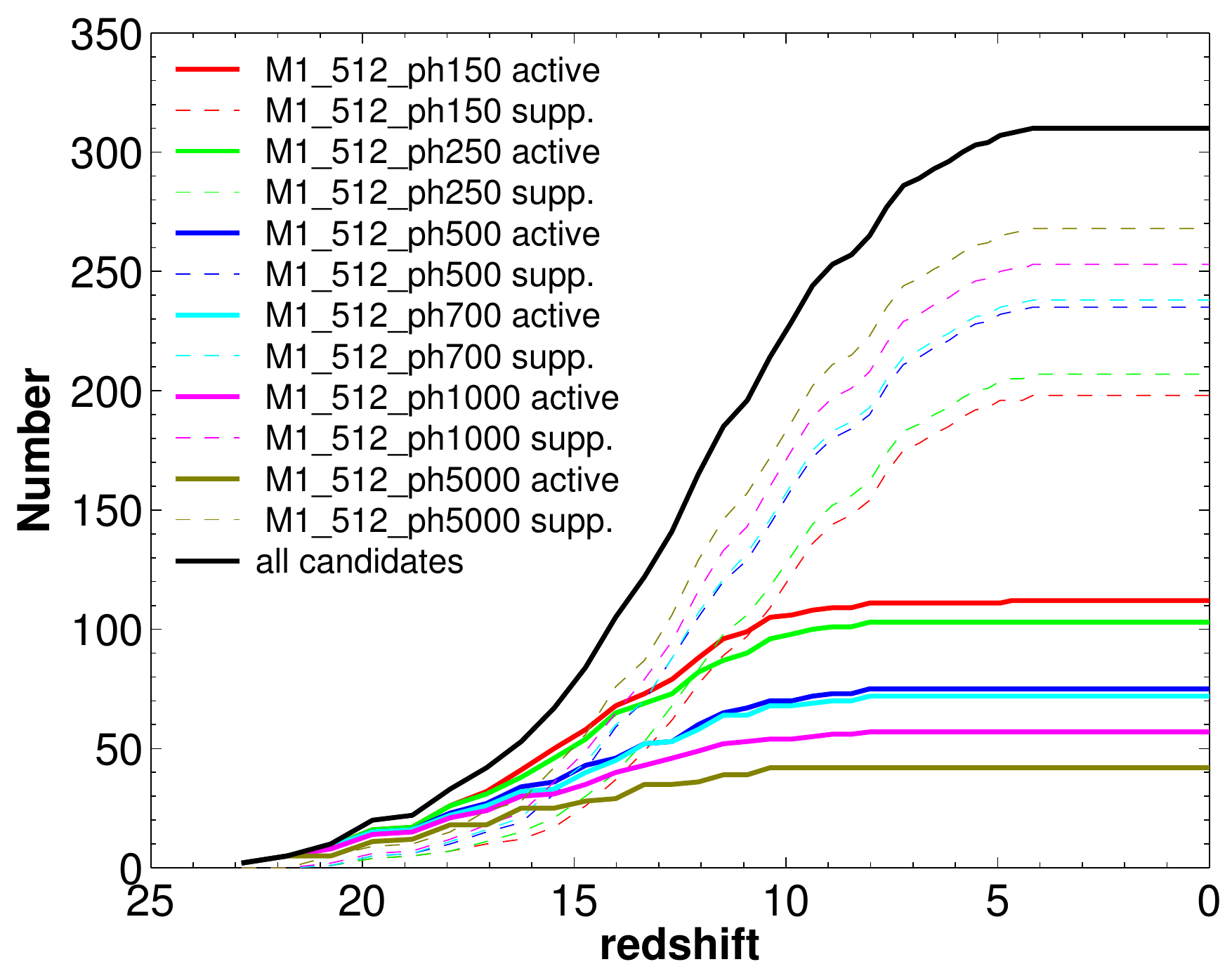}
\caption{Cumulative formation history for each of the respective ionization efficiencies. Solid lines represent active sources and dashed lines represent suppressed sources. The solid black line represents the formation of all potential cluster candidates identified within the Aq-A2 simulation.}
\label{fig:cum_formation}
\end{figure}

Figure \ref{fig:cum_formation} shows the cumulative total of GCs forming indicating a factor of $\sim$3 difference in total number from the lowest ($f_\gamma = 150$ photons/baryon) to the highest ($f_\gamma = 5000$ photons/baryon) ionization efficiency.

Table \ref{tbl:form_num} shows the total number of objects formed and suppressed over the entire simulation for different photo-ionization efficiencies. As expected, there is a drastic dependence of the suppression rate on the ionization efficiency. The number of objects which form between the photon-poor ($f_\gamma = 150$) and photon-rich ($f_\gamma = 5000$) environments in our models differ by a factor of 2.7.


\begin{table}
 \centering
 \small
 \caption{The number of active and suppressed sources of each of the various escape fractions and their corresponding suppression rates. }
 \begin{tabular}{cccc}
\hline
\hline
 Model & $\#$ Active  & $\#$ Suppressed  & Suppression  \\
  Name & GCs & GCs &  Rate [$\%$] \\
\hline
\hline
no suppression & 310  & 0 & 0 \\
M1\_512\_ph150 & 112  & 198 & 0.64 \\
M1\_512\_ph250 & 103  &  207 & 0.67\\
M1\_512\_ph500 & 75  &  235 & 0.76 \\
M1\_512\_ph700 & 72  &  238 & 0.77\\
M1\_512\_ph1000 & 57  &  253 & 0.82\\
M1\_512\_ph5000 & 42 &  268 & 0.86 \\
\hline
\hline
\end{tabular}
\label{tbl:form_num}
\end{table}

\subsubsection{Ages}
As illustrated by the open bins in Figure \ref{fig:formation_hist}, the identification of GC candidates extends down to z = 4. Through the process of self-ionization, the potential number of GCs drastically decreases due to the ionization flux from the first group of active sources. This process pushes back the average age of each of the GC populations. The mean age (calculated assuming a flat cosmology) for Model 1 is shown in Table \ref{tbl:form_ages}.  Overall, a higher the ionization efficiency will lead to, on average, older GCs due to the wiping out of potential sources at low redshift. The youngest population on average is the population resulting from the M$1\_512 \_\mathrm{ph}150$ model, having an average age of 13.30 Gyrs whilst the oldest population on average is the M$1\_512 \_\mathrm{ph}150$ model, having an average age of 13.42 Gyrs. Whilst each of the model GC population's formation is extended in redshift-space, the corresponding age range they occupy is quite narrow (mean =  13.39 Gyr, $\sigma$ = 0.04 Gyr). The surviving GCs in all of the models have mean ages consistent with the Galactic MPGC ages determined from the Advanced Camera for Surveys (ACS) survey carried out by \cite{Marin09} (13.5 $\pm$ 1.5 Gyrs) and other MPGC age studies ($z >$ 5, \citealt{Fan06}).



\begin{table}
 \centering
 \small
 \caption{Mean formation redshift and percentile ages for our metal-poor GCs candidates in each model (assuming a flat cosmology). }
 \begin{tabular}{ccccc}
\hline
\hline
Model & $z_{\mathrm{mean}}$  & \multicolumn{3}{c}{Percentile Age [Gyrs]} \\
    Name &  Formation &  10$\%$ &  50$\%$  &  90$\%$   \\
\hline
\hline
no suppression & 12.25 & 12.82 & 13.20  & 13.36\\

M1\_512\_ph150   &  12.92 & 12.87 & 13.23  & 13.36\\

M1\_512\_ph250   &  13.31 & 12.92  & 13.25 & 13.38 \\

M1\_512\_ph500   &  13.56 & 12.94 & 13.25  & 13.39\\

M1\_512\_ph700  &  13.75 & 13.01 & 13.25 & 13.39\\

M1\_512\_ph1000  &  13.92 & 13.01 & 13.27 & 13.39\\

M1\_512\_ph5000  &  14.05   & 13.01 & 13.27 & 13.39\\
\hline
\hline
\end{tabular}
\label{tbl:form_ages}
\end{table}

\subsubsection{Spatial Distribution}
\begin{figure}
\centering
\includegraphics[width = 0.45\textwidth]{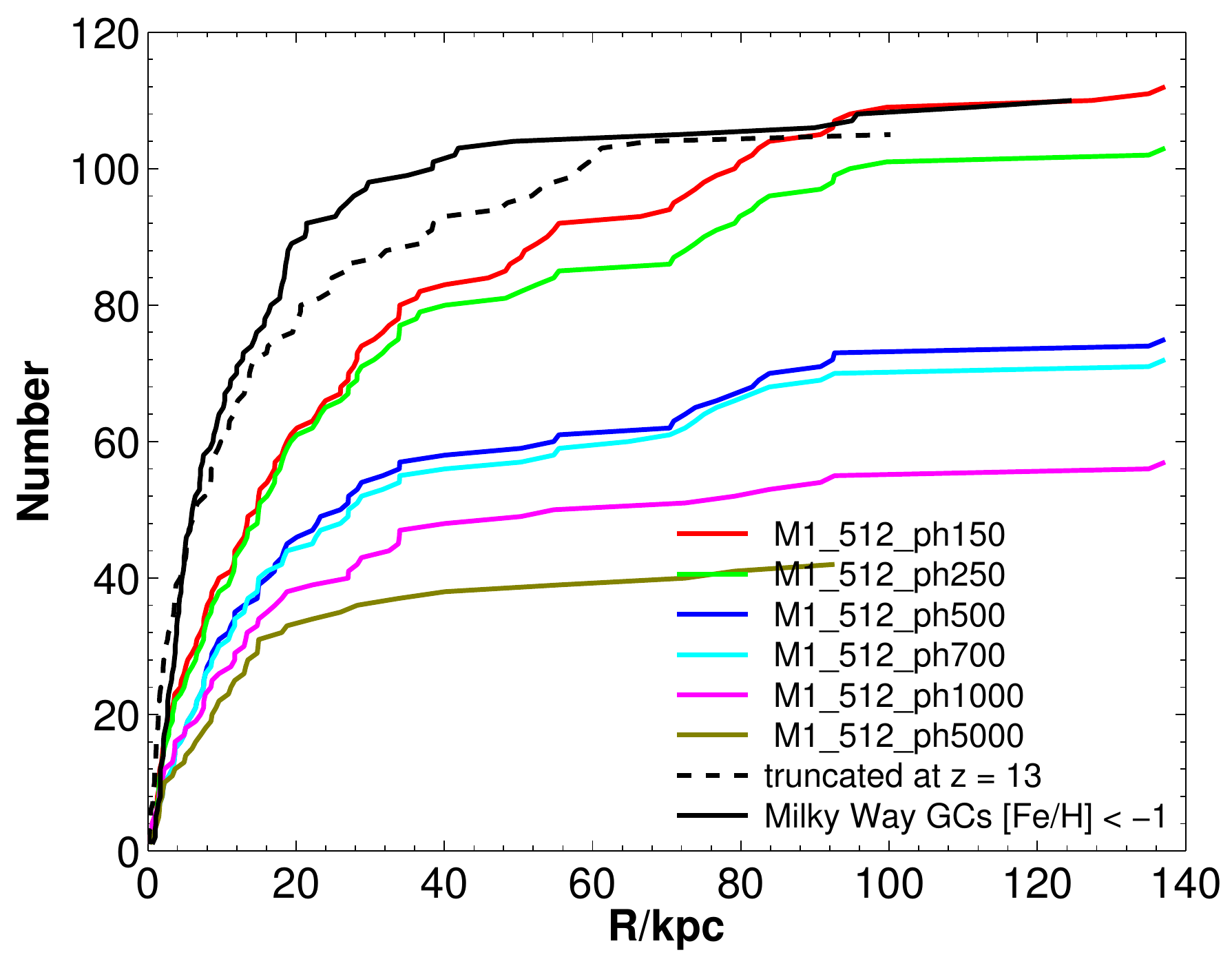}
\includegraphics[width = 0.45\textwidth]{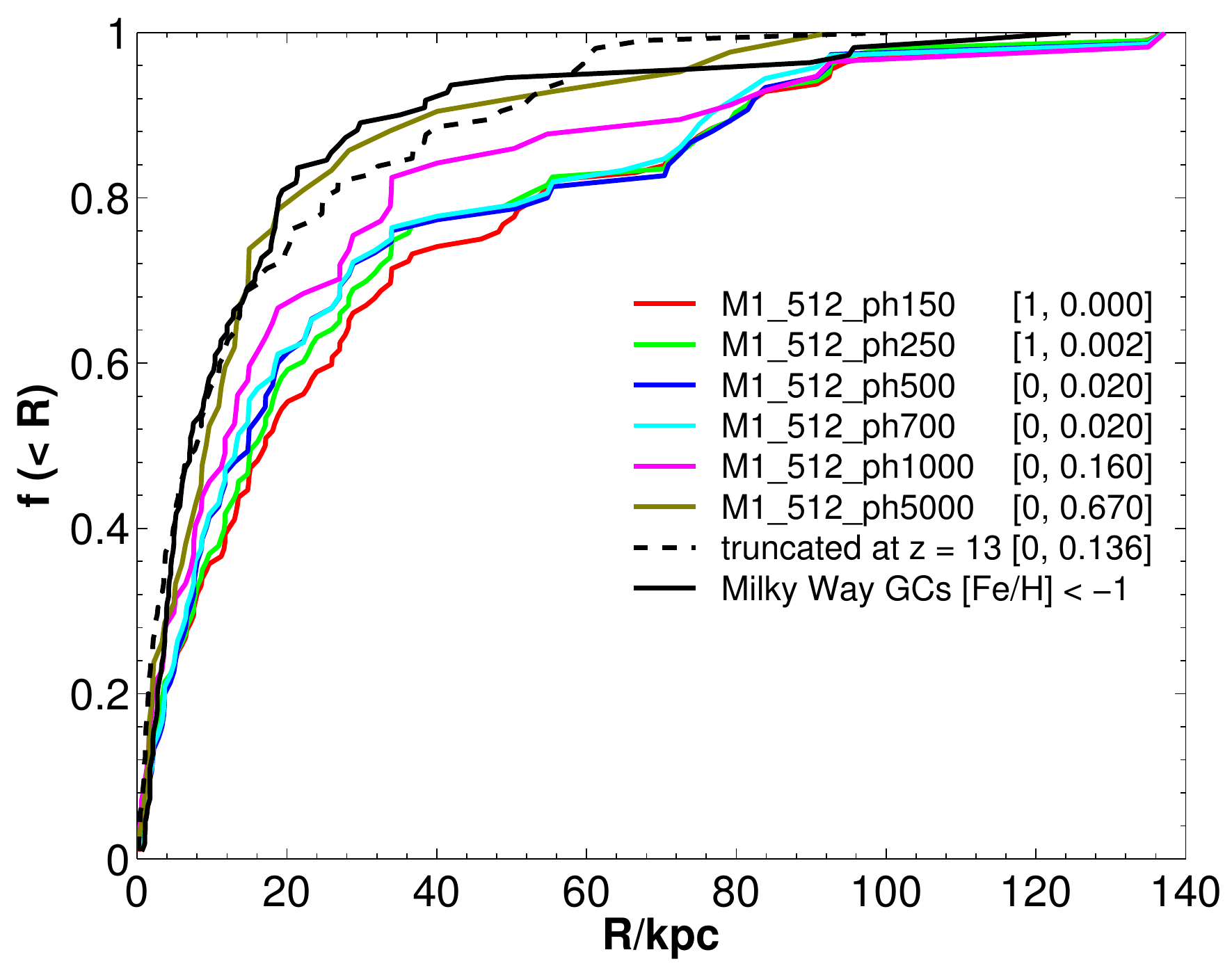}
\caption{Radial distribution of candidate GCs in both raw number and normalised to the total number. The thick black line represents the Milky Way's metal-poor GC distribution. In addition to the six ionization models, the black dashed line represents the truncation model (i.e. GCs in the sample have formed before z = 13 and have merged with the central halo by z = 0). The legend also shows the results from a two-sampled Kolmogorov-Smirnov test carried out between the Milky Way metal-poor distribution and each of the respective models.  Each number corresponds to [H, p] where H = 1 if the test rejects the null hypothesis that they are drawn from the same distribution at the 1$\%$ significance level and $p$ is the asymptotic $p$-value.}
\label{fig:frac_rad}
\end{figure}

Figure \ref{fig:frac_rad} shows the z = 0 radial distributions of the GCs in our simulations in both raw number and as a fraction of the total number. These were obtained by following the most-bound particle of each of the candidates once they had merged with the central halo through to the present day (see Section \ref{subsec:dyndest} for details). The most immediate observation of the is the lack of surviving sources in the inner 50 kpc. Even for the most conservative reionization model (M$1\_512 \_\mathrm{ph}150$), there are $\sim$20 too few objects to reconcile the radial distribution of the AqA2 halo's MPGC distribution with that of the Milky Way's. 

Another significant feature is the difference in shape of the radial distributions of the inhomogenous ionization models (i.e. models using {\sc C$^2$-Ray}) and the truncation model (i.e. using $z_{trunc}$). The truncation model's additional GC numbers  in the central region originate from the inclusion of \textit{all} potential sources before z = 13 which are unavailable to the models using {\sc C$^2$-Ray}. As Figure \ref{fig:formation_hist} illustrates, there are a number of potential sources between z = 23 -- 13 which are otherwise suppressed in the inhomogeneous ionization models. These halos, having been included in the candidate sample in the truncation model reveal themselves in the central regions of the spatial distribution at z = 0. This highlights the first, primary result of this work which is that by treating the ionization field inhomogenous manner, it results in significantly different total numbers and  z = 0 radial distributions of potential GCs when compared to a model adopting arbitrary truncation.

A KolmogorovÐ Smirnov (KS) test carried out comparing the six {\sc C$^2$-Ray} distributions against the Milky Way MPGCs ([Fe/H] $<$ -1) found all models except the M$1\_512 \_\mathrm{ph}150$ and M$1\_512 \_\mathrm{ph}250$ models are inconsistent with the Milky Way metal-poor GC  distribution at the 1$\%$ confidence level. It must be restated that none of the models are designed to replicate the distributions of the Milky Way and these results only show these two of the six distributions (photon-rich models) are statistically consistent with the Milky Way's MPGC system.

It is possible that since we are only using one Aquarius halo that our conclusions are susceptible to scatter. \cite{Lunnan12} found that the halo-halo scatter from the six Aquarius halos is up to a factor of 2--3. This is an important caveat of this work and means that in order to get a proper handle on scatter of this distribution, more Milky Way-type galaxies need to be examined. Further discussion of this scatter is discussed in Section \ref{sec:discussion}.

The time scale for dynamical friction (\citealt{Chandrasekhar43}) is inversely proportional to mass, so the GCs will be more affected by this process than the dark matter particles (which are of order 100 times less massive). For GCs of mass a few times $10^5 M_\odot$ at radii of 10 kpc the dynamical friction time scale is $10^{12}$ years, so this is unlikely to affect our results.

\subsubsection{Contribution To ionization Of IGM}
\begin{figure*}
\centering
\includegraphics[width = 1\textwidth]{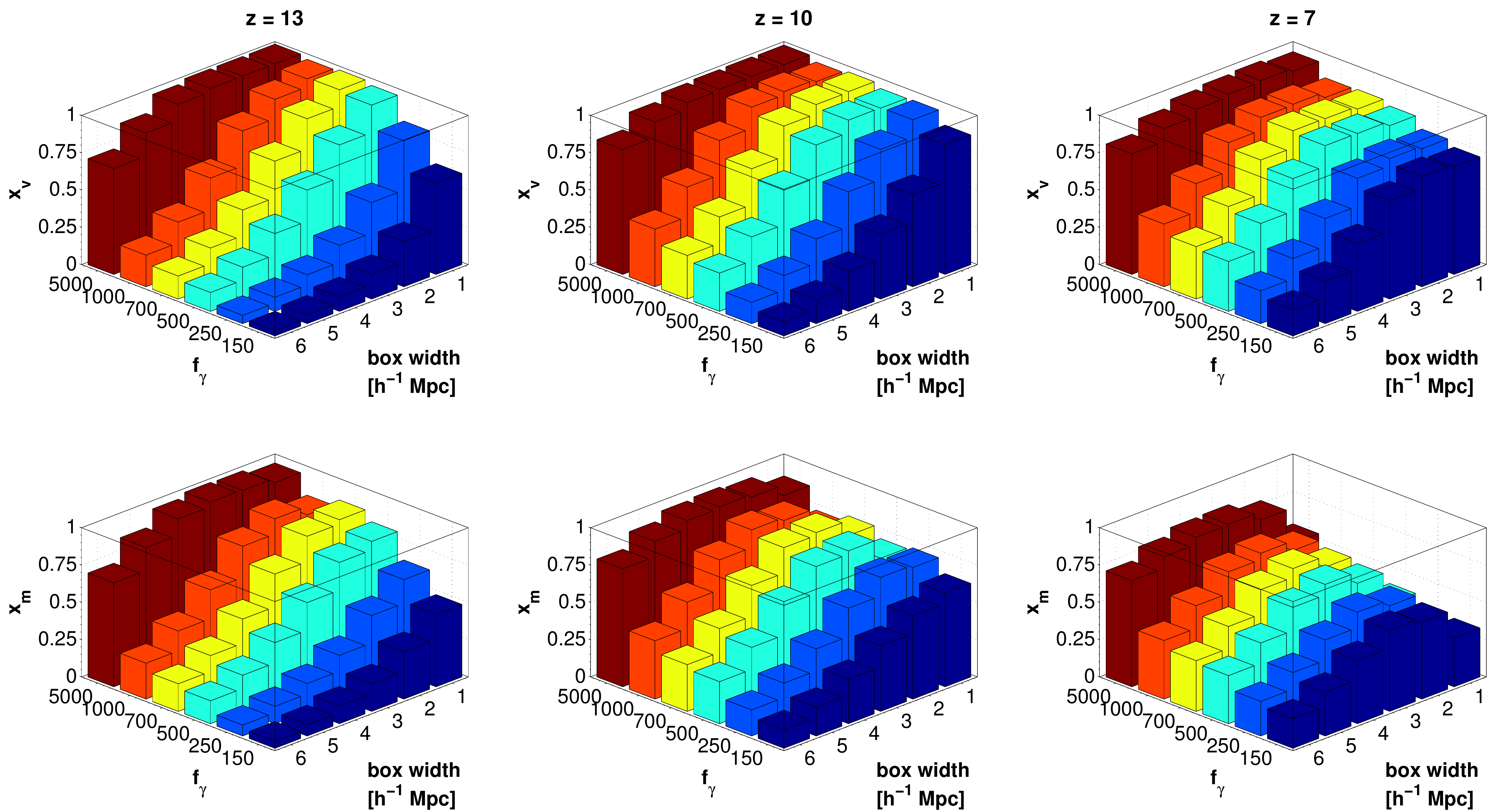}
\caption{The fraction of the volume ($x_v$) and mass ($x_m$) ionized at three different redshifts (z = 7, 10 and 13) plotted simultaneously against the six different photo-ionization efficiencies ($f_\gamma$) and the width of the box surrounding the host halo in comoving coordinates. For each of the models,  at a given redshift, a box was drawn around the central host galaxy (to within the nearest cell) and the volume and mass fraction ionized was calculated. 
Clearly, if MPGCs do form via the dark halo formation channel then their contributions to the ionization of their formation environments is substantial. Interestingly, at low box widths, the ionized mass fraction drops. This is because the number of baryons drastically increases towards the central regions and so the ionizing photons can't penetrate the inner regions of the halo. As discussed in Sub-section \ref{subsec:mapping}, the source sample could be  incomplete at distances greater than 2 $h^{-1}$ Mpc from the host (i.e. box width $>$ 4 $h^{-1}$ Mpc) due to the fact the the high resolution volume from which the candidates are identified is smaller than the total box volume of {\sc{C$^2$-Ray}} (the density field is still accurately represented at all scales however). }
\label{fig:3D}
\end{figure*}

Figure \ref{fig:3D} shows the enclosed mass and volume ionized of each of the six models within a comoving box width (centred on the host galaxy) spanning 1 -- 6 $h^{-1}$ Mpc. Within the largest box size, the amount of volume and mass ionized is small compared to the amount of volume and mass ionized in a small box. This is because candidate formation density is largest in the central most dense region of the simulation. Since a large box width encloses significantly more baryons than a small box, the total volume and mass fraction ionized is lower. As the box width decreases the amount of volume and mass ionized increases as expected. Similarly as the photo-ionization efficiency ($f_\gamma$) increases, the fraction of volume and mass of the respective enclosed volume also increases as expected. In the extremely photon rich model (M$1\_512 \_\mathrm{ph}5000$), as much as 80$\%$ of the entire simulated is ionized by z = 10.

In terms of volume, 60$\%$ and 98$\%$ of the 2$^3$ $h^{-3}$ Mpc$^3$ volume centred on the AqA halo is fully ionized by z = 10 for the photon-poorest (M$1\_512 \_\mathrm{ph}150$) and photon-richest (M$1\_512 \_\mathrm{ph}5000$) models, respectively. In terms of the total mass ionized, 50$\%$ and 90$\%$ of the total mass within the same volume was ionized by z = 10 for the photon-poor and photon-rich models, respectively. The subsequent decrease in the ionized mass fraction is due to the material recombining at low redshift. The ionized volume however, remains relatively constant beyond z = 8 since there are fewer sources below this time to increase the ionized volume fraction. 

The number of ionizing photons emitted by all active sources is plotted in Figure \ref{fig:cumulativephotons}. The most efficient of our models peaks at 10$^{70}$ photons being produced over the course of the simulation. The active sources in our photon poorest model produces approximately 10$^{68.5}$ photons within the total 6$^3$ $h^{-3}$ Mpc$^3$ volume.

\begin{figure}
\centering
\includegraphics[width = 0.45\textwidth]{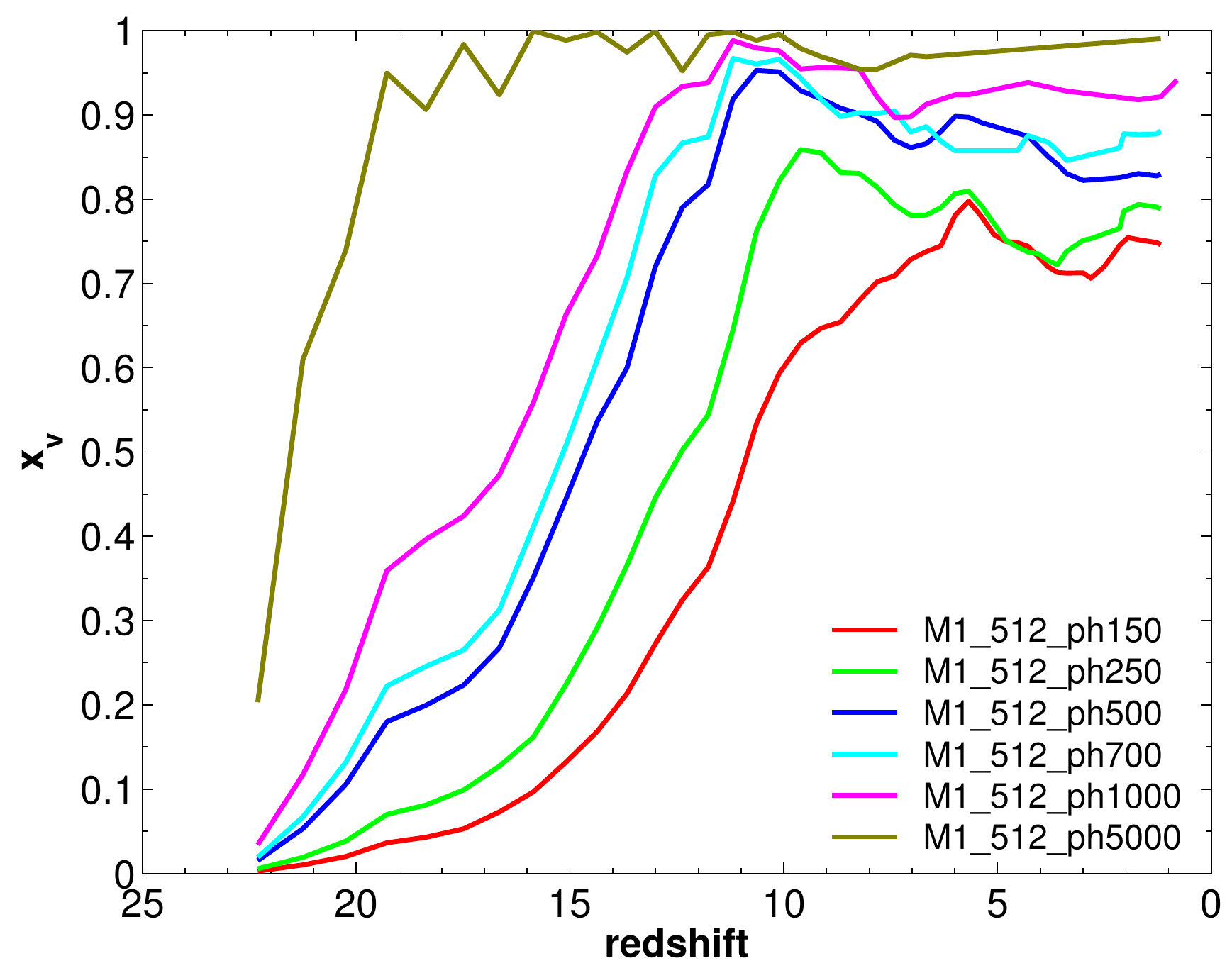}
\includegraphics[width = 0.45\textwidth]{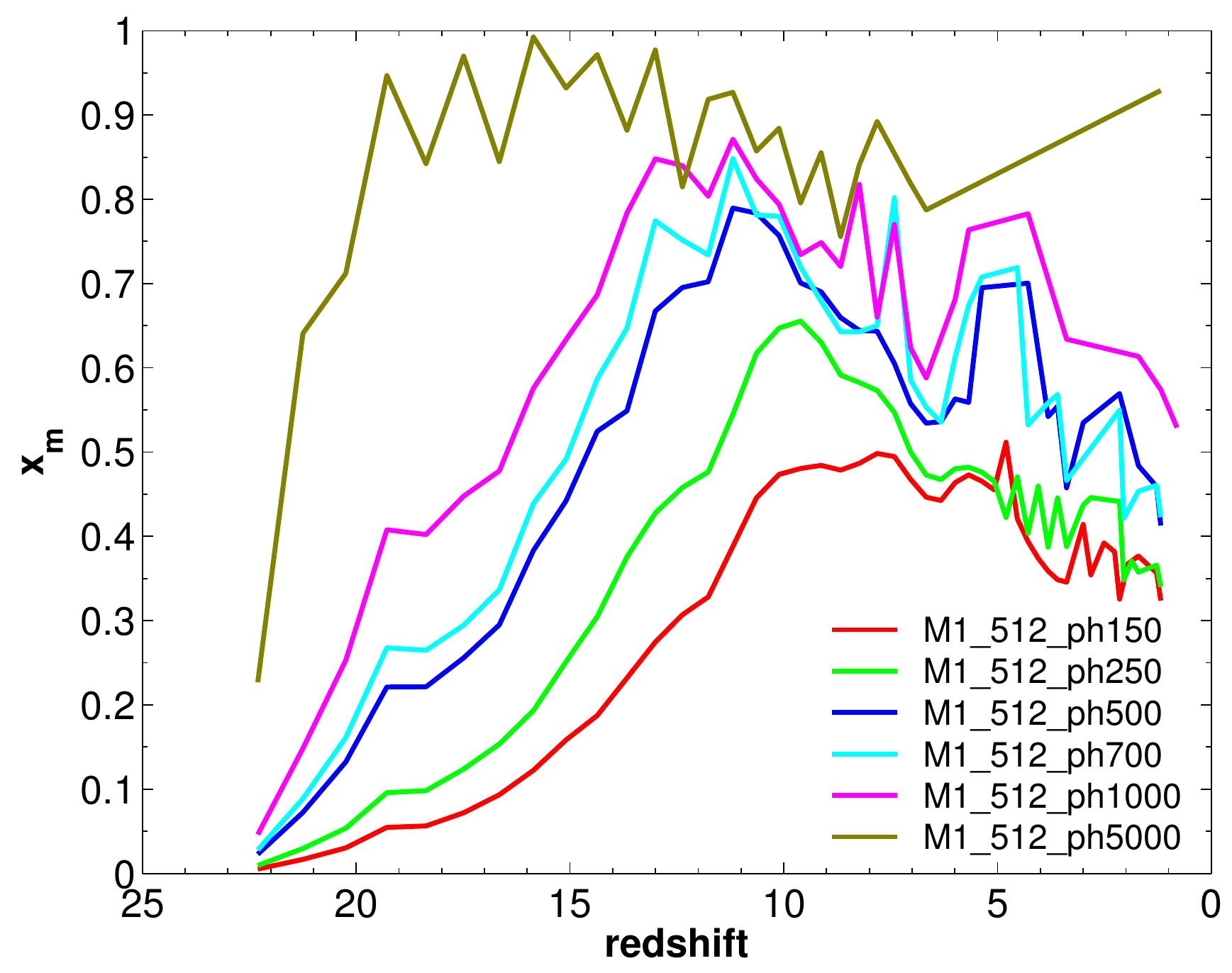}
\caption{Redshift evolution of the volume ($x_v$) and mass ($x_m$) fraction ionized within a $2^3$ $h^{-3}$ Mpc$^3$ box centred on the host galaxy for each of the respective photo-ionization efficiencies. Whilst a significant quantity of the enclosed volume is ionized across each of the models, recombinations at low redshift decrease the mass fraction ionized since the number of active sources below z $<$ 10 drastically reduces. In the late reionization era, overlap by neighbouring ionizing sources would ensure the remaining neutral gas becomes ionized by z = 0 (see \citealt{Iliev11}).}
\label{fig:vol_ion}
\end{figure}

\begin{figure}
\centering
\includegraphics[width = 0.45\textwidth]{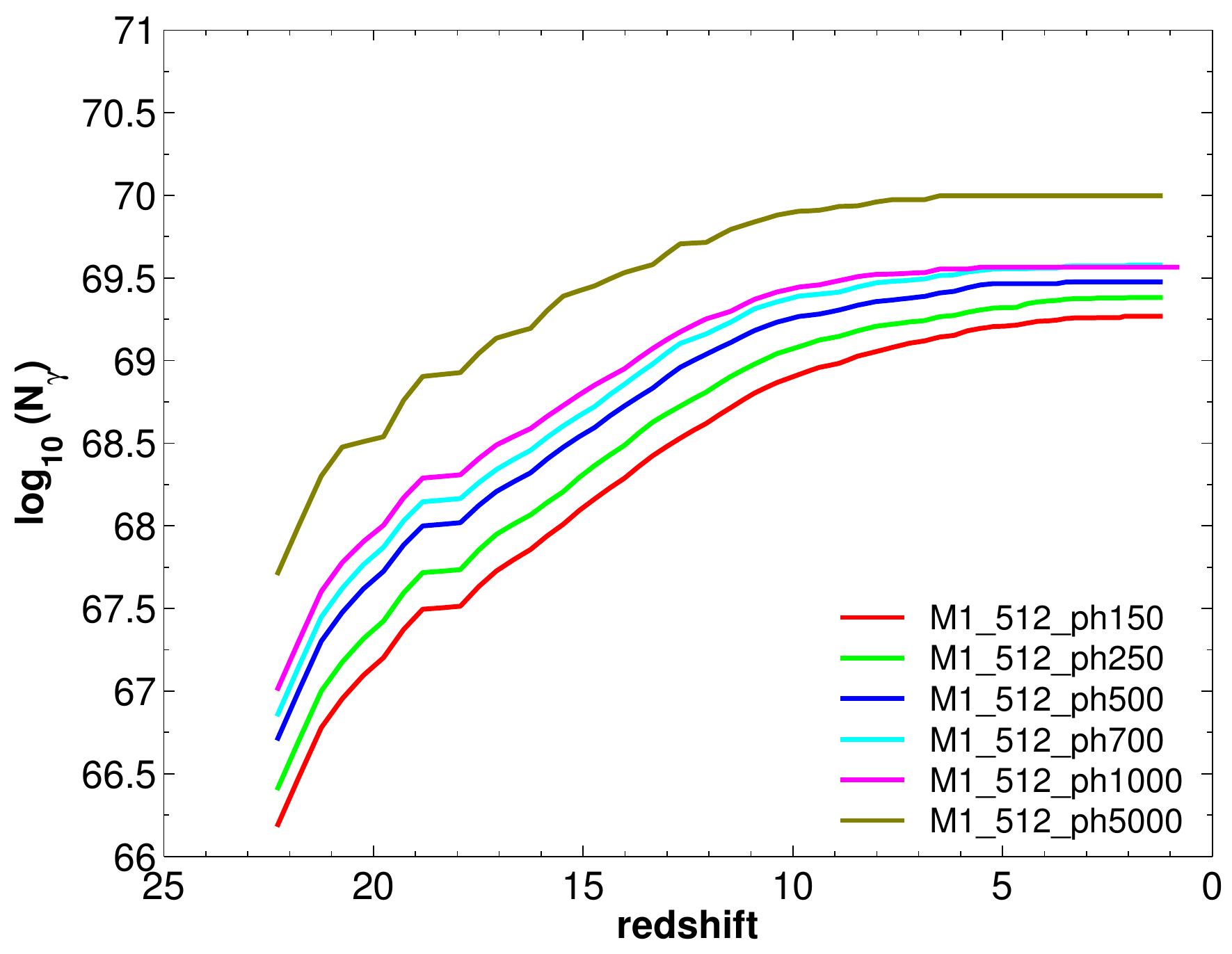}
\caption{Cumulative number of ionizing photons emitted by MPGC candidates within Aquarius over the course of the simulation. }
\label{fig:cumulativephotons}
\end{figure}

The primary result from this analysis is that if MPGCs did form via the mechanism tested, then at worst, MPGCs contributed a \textit{non-trivial} fraction ($\ge$ 40$\%$, Figure \ref{fig:vol_ion}) of the total ionization of the IGM within a distance of $\sim$1 $h^{-1}$ Mpc of the Milky Way

\subsection{Extensions to the Model: Dynamical Disruption $\&$ Delayed GC Formation}
In this subsection we discuss extensions to the model. In order to fully appreciate the effect dynamical disruption and delayed formation have on cluster properties, we present our results separately. 

\subsubsection{Dynamical Disruption}

In order to better understand the survival rates of MPGCs once they merge into the central host halo, we adopted the dynamical disruption model of \cite{Baumgardt03}. The details of the method of how we include dynamical destruction are discussed in Sub-section \ref{subsec:dyndest}. Using this method, the dissolution times for all of the candidate clusters were calculated.  Objects which have a dissolution time less than the look-back time to when the merged with the host are disrupted. The accuracy of this type of modelling has not yet been compared with higher resolution analyses of small scale hydrodynamic systems and so we present these results well aware of the several caveats involved (see Section \ref{subsec:caveats} for details) and use this preliminary work as a  launching platform for future research.

\begin{figure}
\centering
\begin{tabular}{c}
\includegraphics[width = 0.45\textwidth]{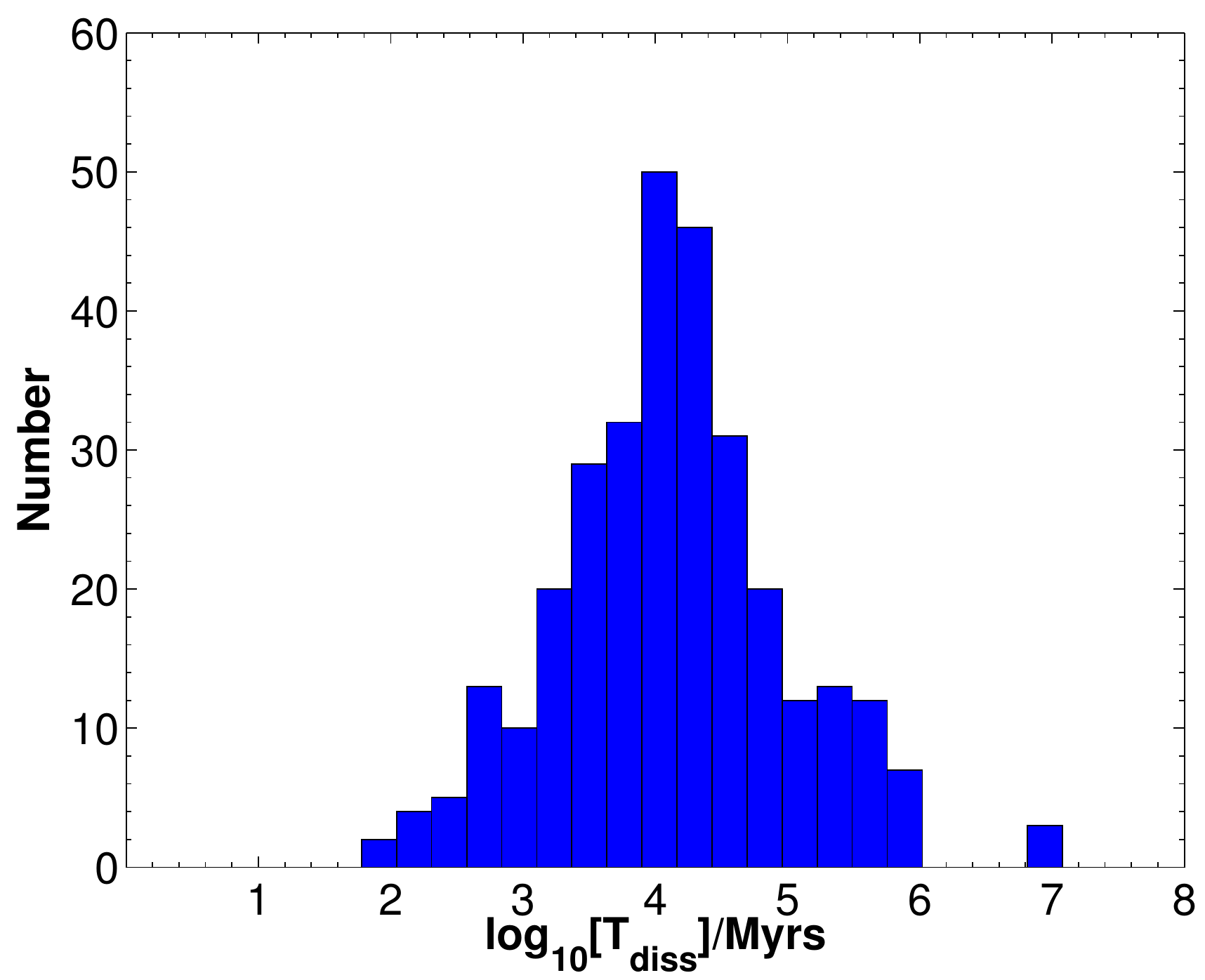}
\end{tabular}
\caption{Dissolution time of candidate MPGCs which have merged with the central host by z = 0. An object with a dissolution time less than the look-back time it merged with the central host are disrupted.}
\label{fig:Tdiss}
\end{figure}

The sample includes all candidate clusters identified via the temperature-velocity dispersion threshold discussed in Sub-section \ref{GCformmodel}. As shown in Figure \ref{fig:Tdiss}, just over half (60$\%$) of the clusters identified, whether they are suppressed or not, will undergo dynamical disruption. Further analysis of different King profiles indicated only a marginal ($\pm$$6\%$) swing in survival probability. 

\begin{figure}
\centering
\begin{tabular}{c}
\includegraphics[width = 0.45\textwidth]{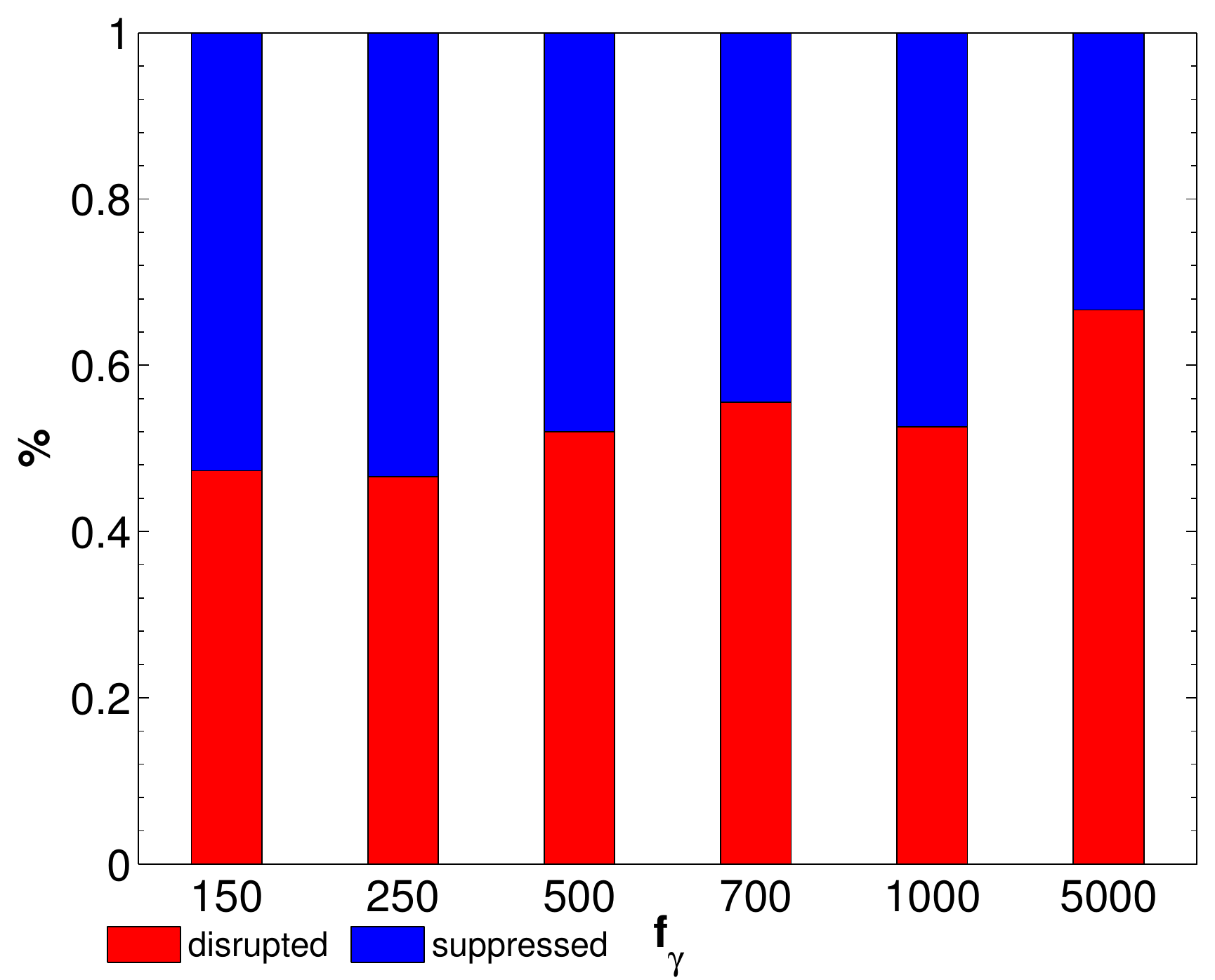}
\end{tabular}
\caption{Accounting for the empty or disrupted halos. Each stacked bar represents the percentage fraction lost due to suppression and dynamical disruption. With the exception of the photon-richest model ($f_\gamma = 5000$), approximately half are consistently lost due to disruption and just under half are lost due to suppression in each of the respective models.}
\label{fig:supsurv}
\end{figure}

Figure \ref{fig:supsurv} presents a percentage break down of what happens to the sources that don't survive through to the present day. The only two ways of removing sources in this model is by dynamical disruption or suppression by ionization. With the exception of the photon-richest model ($f_\gamma = 5000$), approximately half of the candidate GCs in each model are lost due to dynamical disruption and just under half are lost due to suppression via photo-ionization. The reason the suppressed fraction doesn't increase more significantly as the ionization efficiency increases is because the sample only examines the sub-population of clusters which are \textit{destroyed} (i.e. all candidates of the initial population which were removed via suppression or disruption). For example, in the less efficient models, more clusters are going to be included in the sample because fewer clusters are suppressed at lower redshift. This also means however that a larger percentage of these clusters are susceptible to disruption.

\begin{figure}
\centering
\begin{tabular}{c}
\includegraphics[width = 0.45\textwidth]{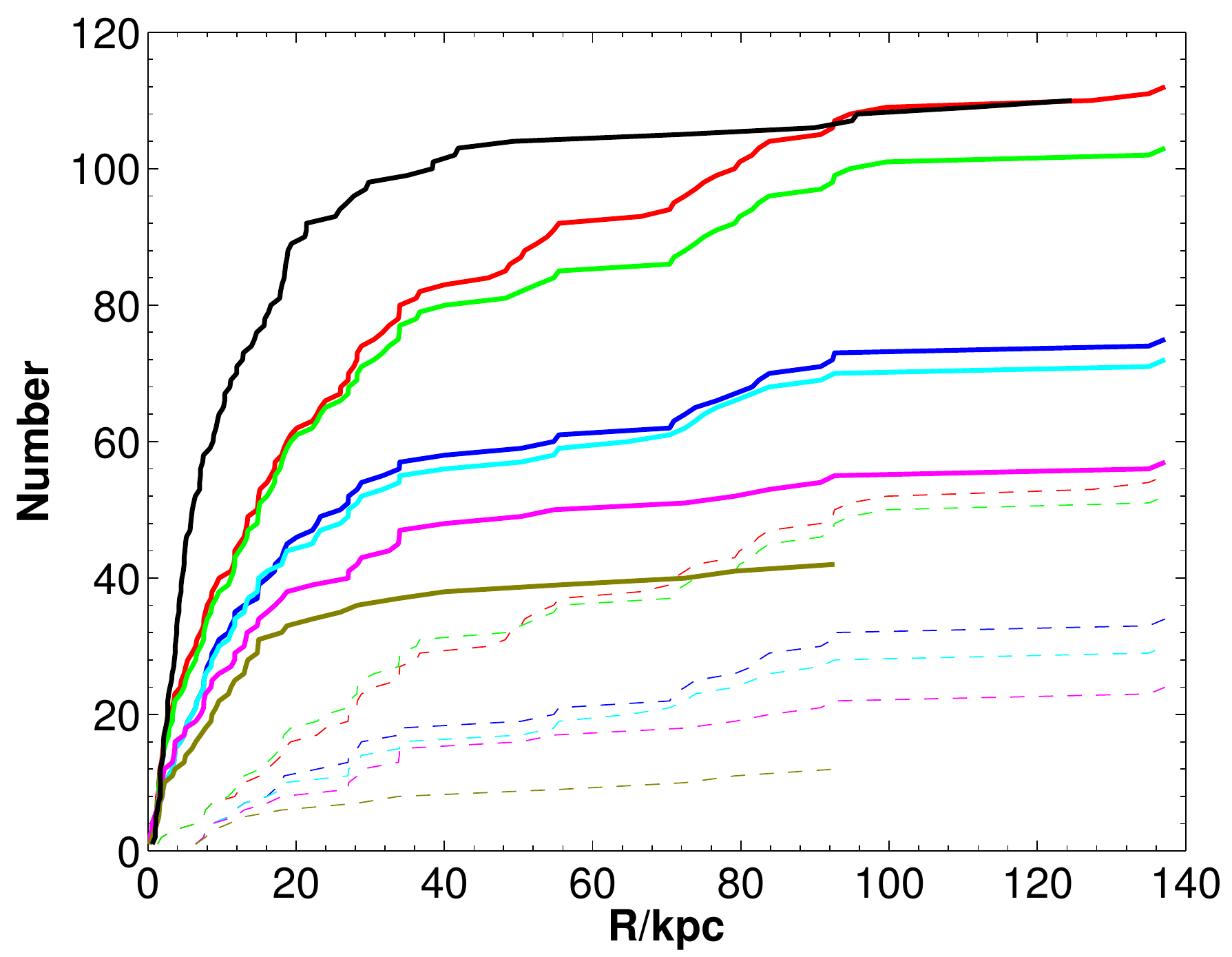} \\
\includegraphics[width = 0.45\textwidth]{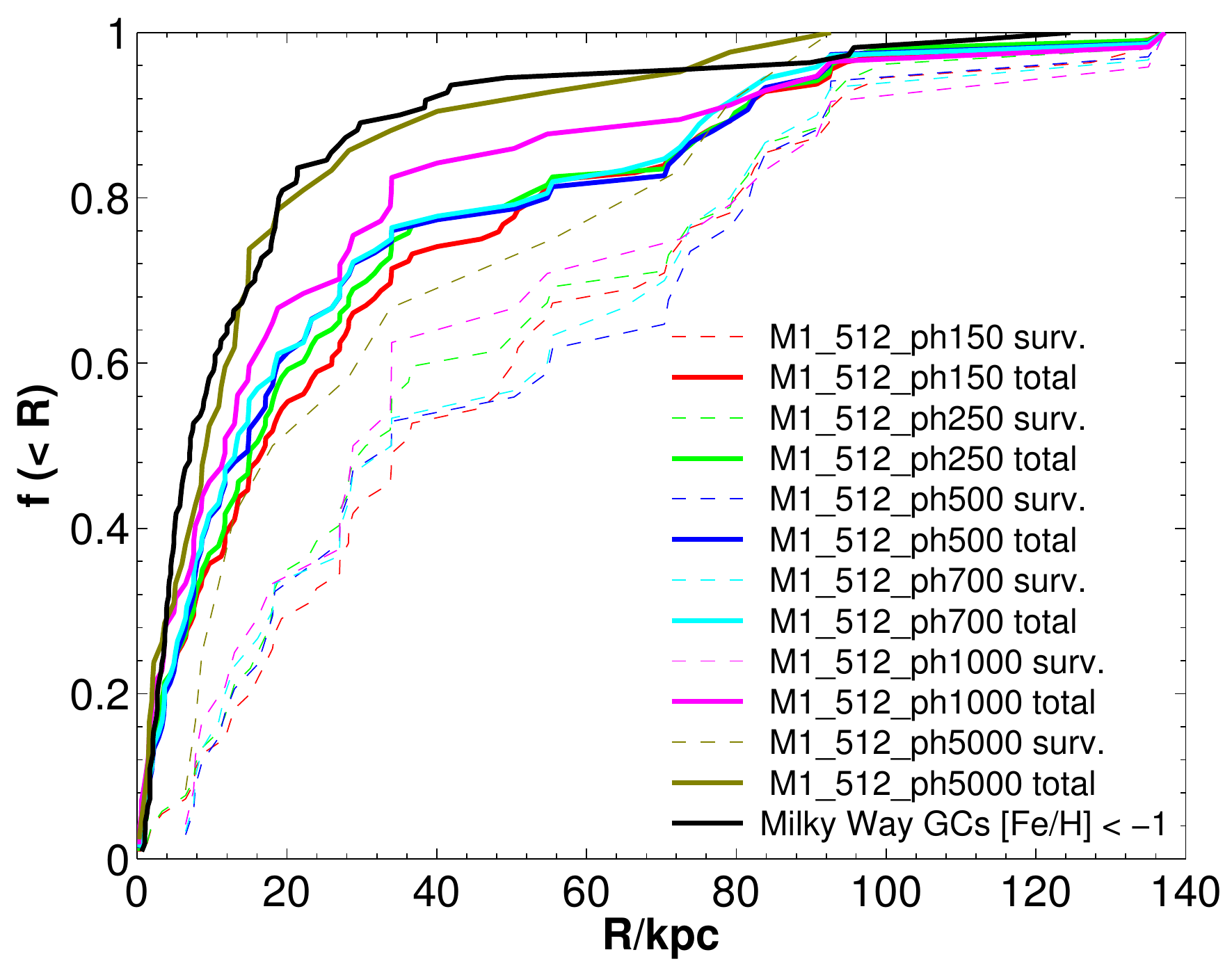}
\end{tabular}
\caption{Model 1 (permanent suppression) radial distributions of candidate MPGCs in both raw number (top panel) and normalised number (bottom panel) for models with (dashed lines) and without (solid lines) dynamical disruption. The effect of the dynamical model is stark. Candidate GC populations in the central regions are drastically reduced resulting in far more extended radial distributions. For reference, the upper panel has the same legend as the lower panel but we exclude it for clarity.}
\label{fig:disruptradial}
\end{figure}

The drastic decrease in the number of potential clusters (excluding the effect of suppression) has significant effects on the spatial distributions of the potential MPGC candidates. Figure \ref{fig:disruptradial} plots the radial distribution of each of the models with and without dynamical disruption. Unlike the original model, none of the disruption models are statistically consistent Milky Way MPGCs at a 1$\%$ significance level. 


In order to gain a more complete understanding of the dynamical processes occurring within the host galaxy after a merger takes place, a full hydrodynamic, high-resolution simulation would be required in order to properly determine the survival rates of GCs. We present these broad brush dynamical destruction results to show in concept, that large quantities of GCs will be destroyed once they merge with the central host galaxy and that their radial distributions are currently irreconcilable with the present day metal-poor Galactic GC population.

\subsubsection{Model $\#2$: Delayed GC Formation}
In the results thus far, we have presented the properties of MPGC candidates in the context of a model which suppresses halos once and permanently. This approximation is accurate so long as there are sufficiently high numbers of ionizing photons available to keep the entire volume ionized. If for instance a previously suppressed halo's descendent's gas content became neutral such that it could  collapse to form a GC, this candidate would be excluded in the previous model. As detailed in Section \ref{subsection:delayedmodel}, we implement a delayed model whereby all suppressed candidates are still potential sources provided they a) enter a neutral region of the IGM and b) have \textit{no} active ancestors. Utilising the merger trees of the AqA2 halo enabled us to implement this model in a relatively straight forward manner.

\begin{figure}
\centering
\begin{tabular}{c}
\includegraphics[width = 0.45\textwidth]{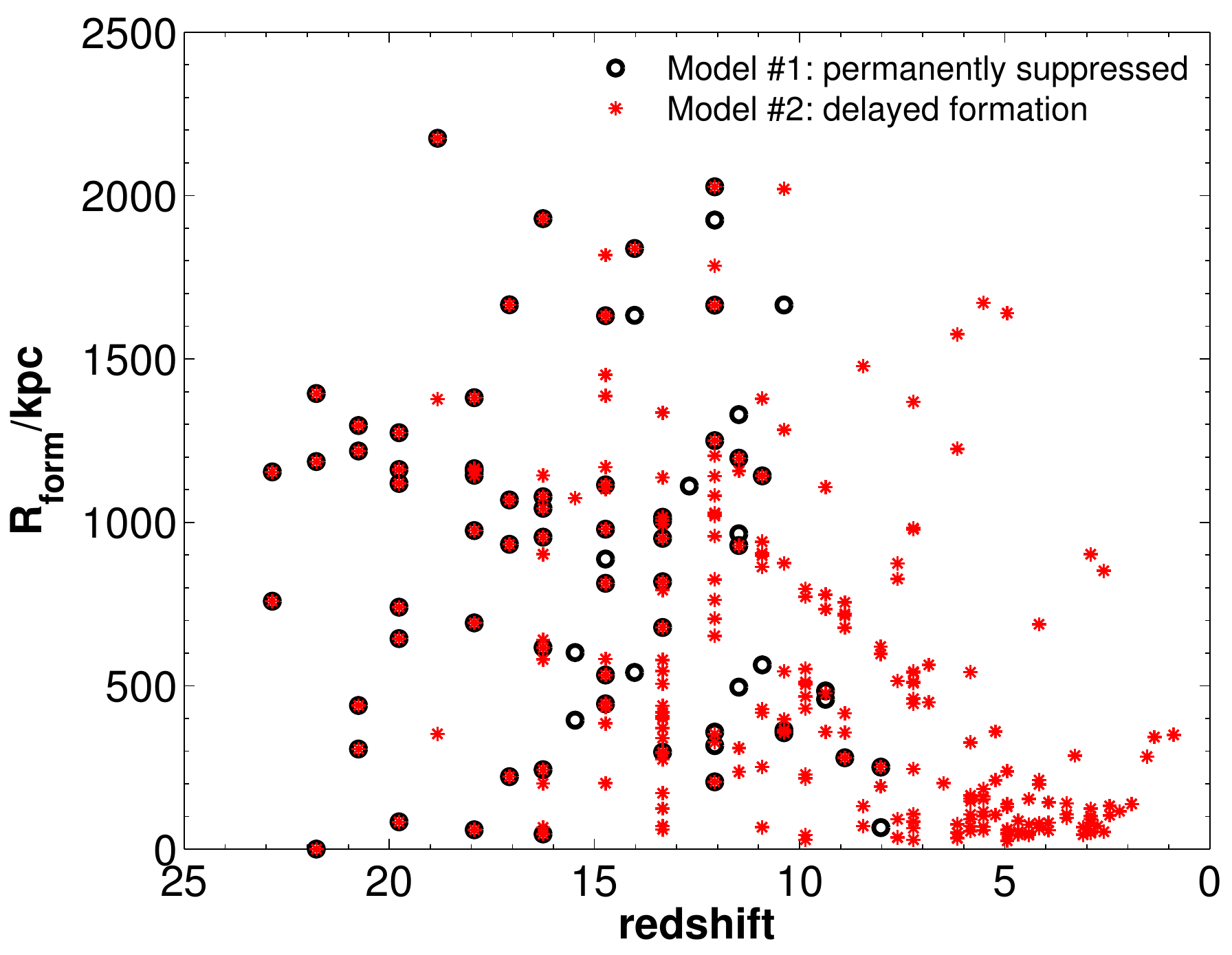} \\
\includegraphics[width = 0.45\textwidth]{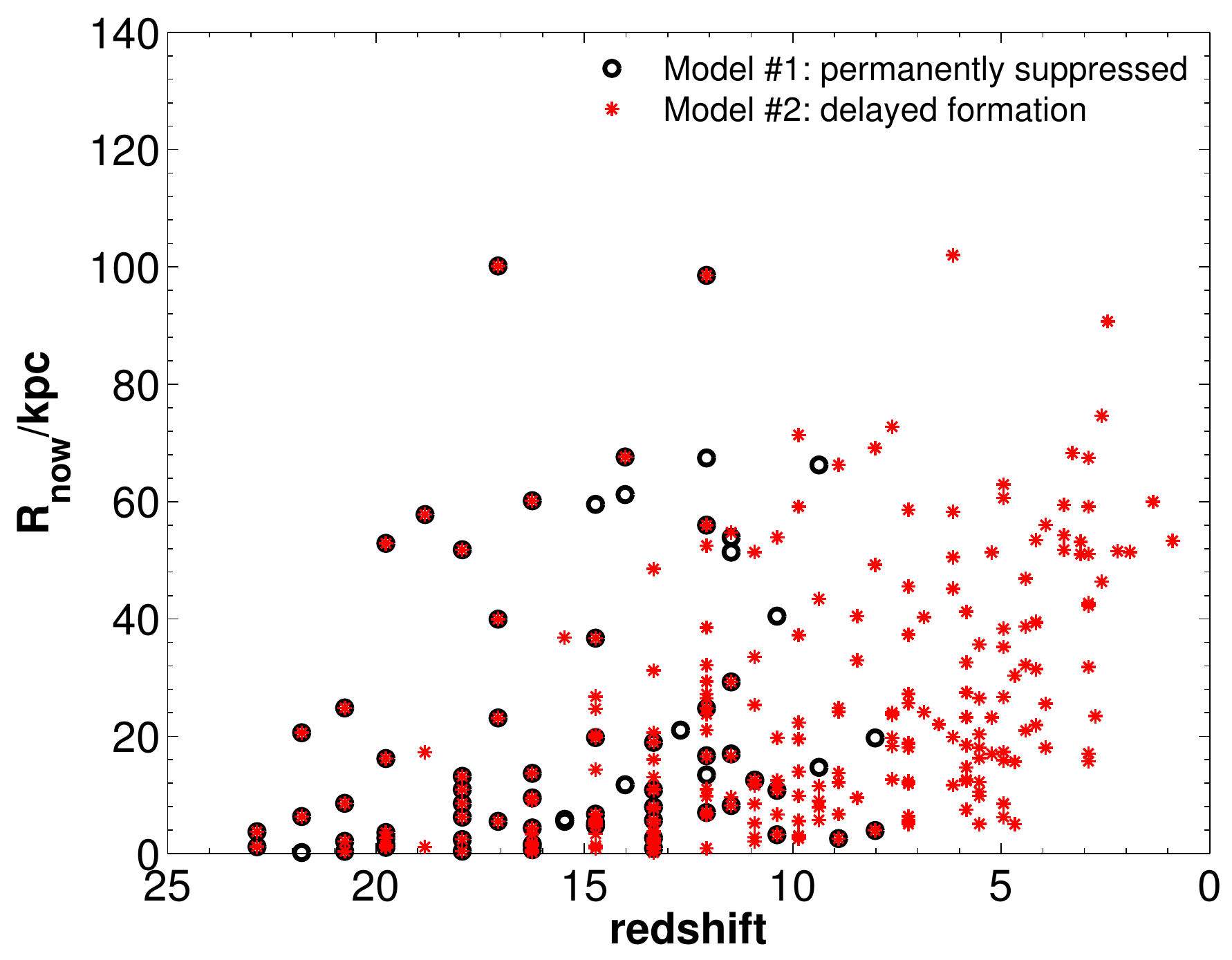}
\end{tabular}
\caption{\textit{Top Panel:} Comparison between Model 1 (permanently suppressed, solid black circles) and Model 2 (delayed formation, red stars) using their respective comoving galactocentric distance at formation and their redshift of formation.  \textit{Bottom Panel:} Similar to the top panel but this time shows the galactocentric distance at z = 0 and redshift of formation of each of the candidates. There are extra sources forming at z $<$ 8  in Model 2 due to previously suppressed sources becoming active after meeting the necessary environmental conditions within R $<$ 300 $h^{-1}$ kpc of the host galaxy.}
\label{fig:permdelay}
\end{figure}

Figure \ref{fig:permdelay} compares the formation history of the MPGC candidates in both the permanently suppressed and delayed formation models in relation to their present day and formation galactocentric radii. At early times, the source positions between the two models are in exact agreement with one another (red stars within black circles) since the IGM hasn't begun to be significantly ionized. As the ionizing flux from those first objects increases over time, the number of suppressed candidates also increases, resulting in a number of extra potential sources which could become active later if the environmental requirements are satisfied. The first of these sources becomes noticeable at z $\sim$ 19 whereby a few previously suppressed halos become active. Due to the non-linear nature of the ionizing field, these first additional sources then lead to more significant changes at later times. The majority of delayed candidates activate at z $<$ 8 within a galactocentric distance of R $<$ 300 $h^{-1}$ kpc. This is a result of there being a large neutral region in the centre due to the recombination rate being larger than the ionization rate. This neutral region can also be reasonably  seen upon a visual inspections of the central region in Figure \ref{fig:diff_fgamma}. It is clear that previously suppressed sources are merging into these regions and subsequently meeting the necessary environmental requirements to become active.

\begin{figure}
\centering
\begin{tabular}{c}
\includegraphics[width = 0.45\textwidth]{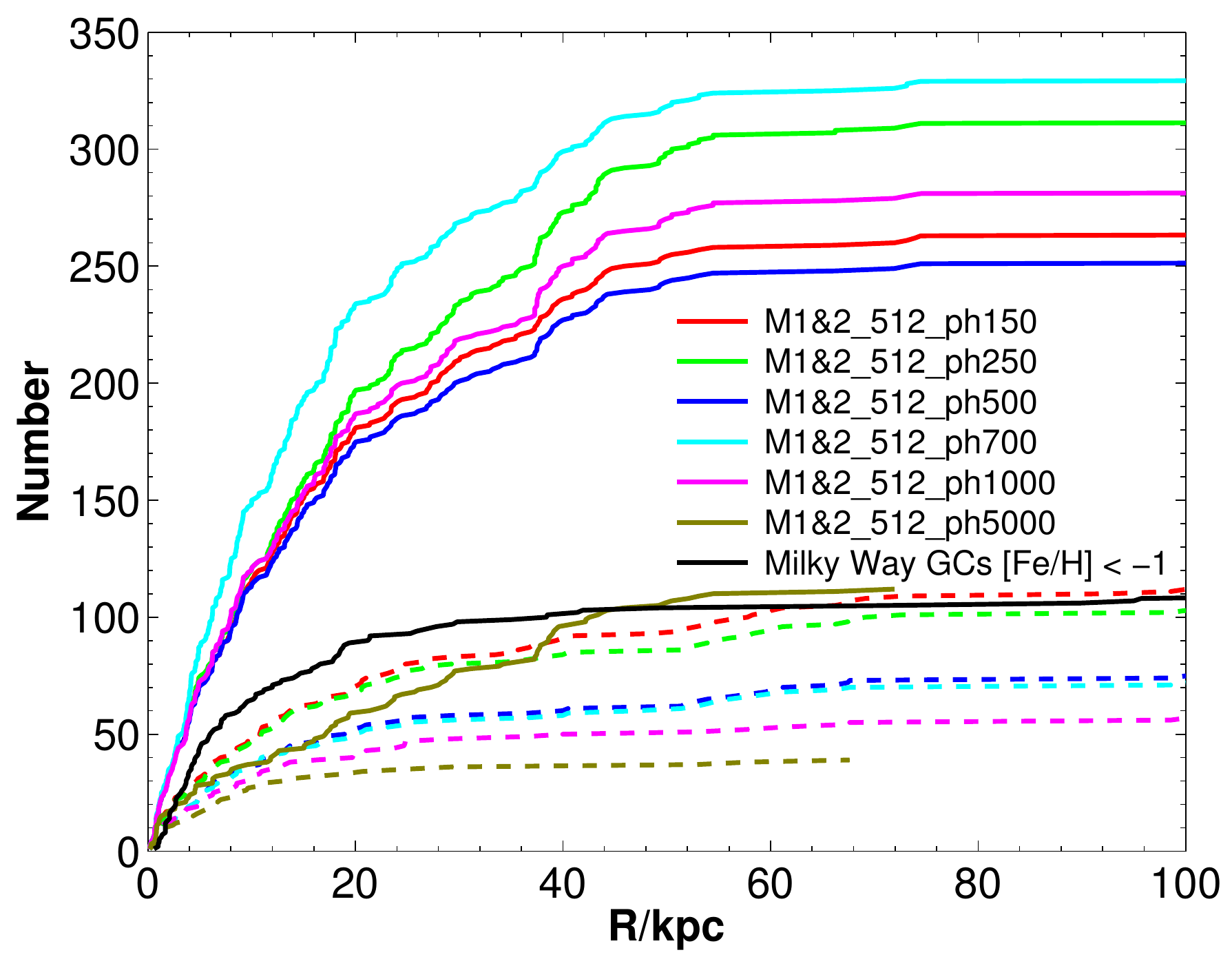}\\
\includegraphics[width = 0.45\textwidth]{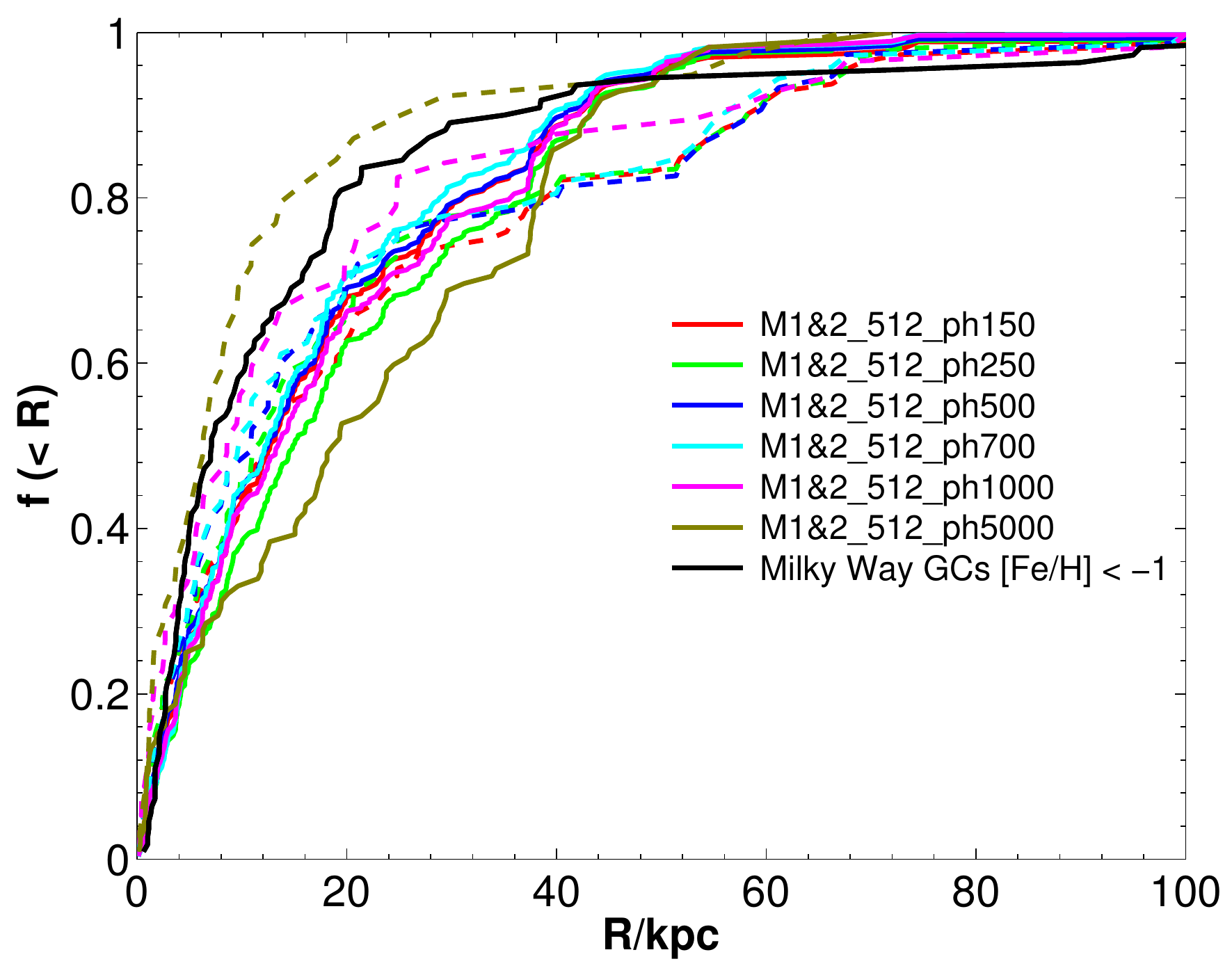}
\end{tabular}
\caption{Cumulative (top panel) and normalised (bottom panel) radial distributions of all photo-ionization efficiencies in both Model 1 (permanently suppressed, dashed) and Model 2 (delayed formation, solid). The delayed model adds as much as three times more potential GC candidates than the equivalent $f_\gamma$ permanently suppressed model (Model 1). }
\label{fig:permdelaydistribution}
\end{figure}

Figure \ref{fig:permdelaydistribution} compares the z = 0 radial distributions of all of the candidate MPGCs (cumulative and normalised) in both the permanently suppressed and delayed formation models. The immediate noticeable difference is the increase in the number of objects in the central region (R $<$ 40 kpc) in the delayed model when compared to Model 1 of the same ionization efficiency. As indicated in Figure \ref{fig:permdelay}, the majority of these new candidates are coming from suppressed sources becoming active in within 300 kpc of the host. These delayed halos subsequently merge with the host galaxy and populate galactocentric distances between 0 and 100 kpc. Despite the extra GCs are small galactocentric radio, a KS-test carried out on all six Model 2 distributions rejects the null hypothesis that they are drawn from the same distribution as the Milky Way MPGCs at the 1$\%$ significance level.

\begin{figure}
\centering
\begin{tabular}{c}
\includegraphics[width = 0.45\textwidth]{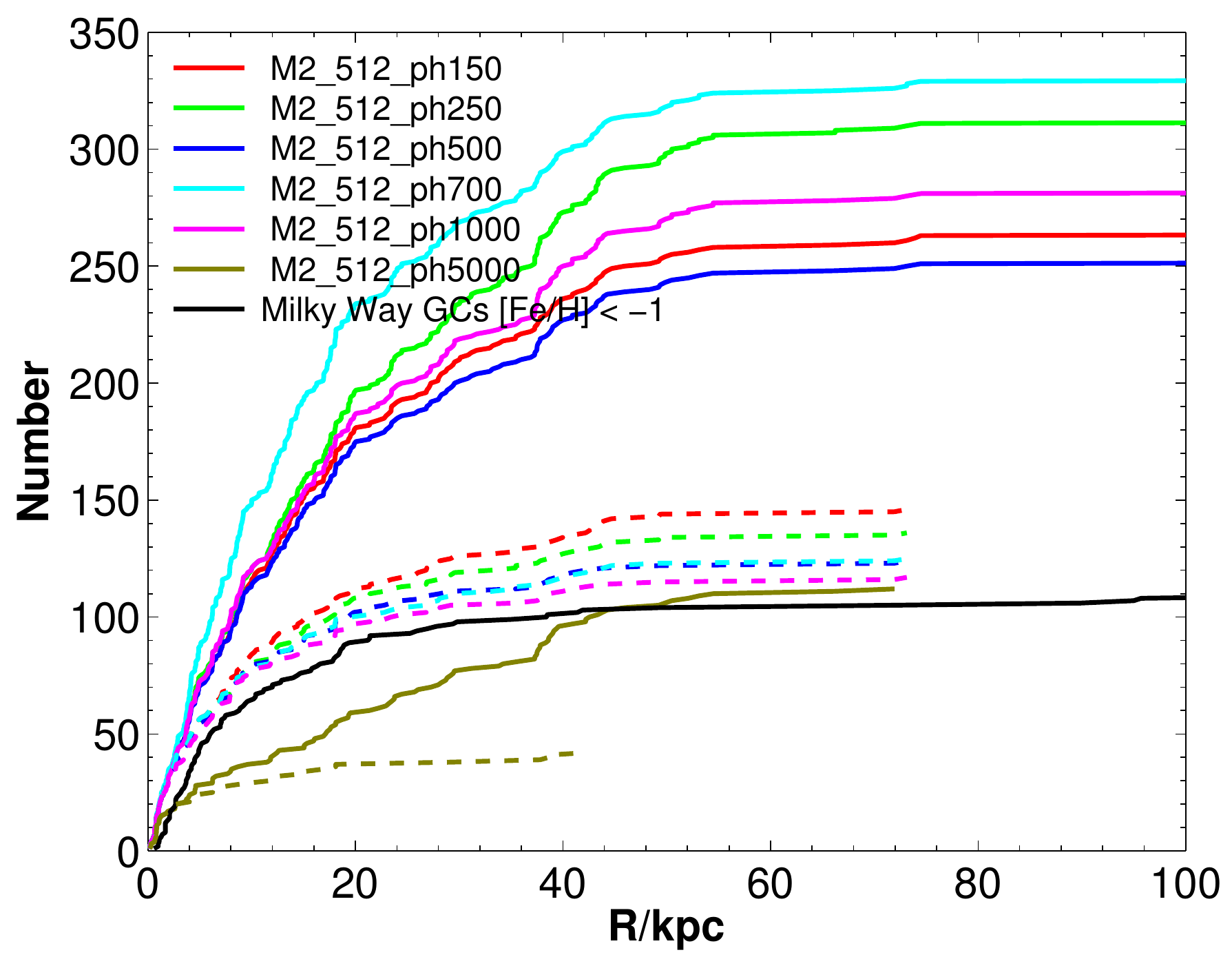}
\end{tabular}
\caption{Radial distributions of all photo-ionization efficiencies in Model 2 (solid) and the same objects truncated at z = 10 (dashed). This shows that a better match to the Milky Way can be obtained if further sources of ionization are added at late redshifts.}
\label{fig:delayedvstrunc}
\end{figure}

Figure \ref{fig:delayedvstrunc} compares the z = 0 radial distributions of all of the candidate MPGCs in the delayed formation models (solid lines) and the same objects truncated at z = 10 (dashed lines). Whilst the delayed model candidates are far too extended in their entirety, when they are truncated their spatial distributions are more closely consistent with the metal-poor GCs of the Milky Way (excluding the M$1\_512 \_\mathrm{ph}5000$). This suggests that if metal-poor GCs did form via the dark halo formation channel, then either external ionization (i.e.\ ionization from the Local Group) or ionization from other unaccounted for sources (e.g. Population II stars) must dominate the local environment from z = 10 onwards in order for their radial distributions to be consistent with that of the Milky Way.

\section{Discussion}
\label{sec:discussion}
In this Section, we discuss the more general achievements and limitations of our models with a particular emphasis on areas of improvement.
\subsection{Overview}
In our previous work (\citealt{Griffen10}) we set out to build a launching platform for more detailed work. This proof-by-concept work proved fruitful in a few important ways. Through a relatively simple GC formation model, we found that if GCs could have formed (in-part) at the centre of dark halo potentials, then the contributions of metal-poor GCs stemming from these dark halos to the reionization of the IGM was significant.  Using the cumulative mass of these primordial GCs, we estimated their ionizing flux and came to the conclusion that it would have been possible for GCs to have ionized the entire Milky Way by z = 13 (for $f_\gamma$ = 210). Despite the limited success of the model, there were a few glaring assumptions or limitations. For example, a) 100$\%$ of the ionizing flux coming from each GC was assumed to have gone into ionizing the Milky Way, b) only one value for the emissivity of the GCs was examined, c) the state of the IGM was entirely unknown at high redshifts and d) an arbitrary truncation redshift based on relatively naive assumptions about the nature of the IGM. This work provided us with the understanding required to develop a more detailed reionization model and to look at ways of improving the treatment of the ionization field.

In this latest work, we improve on all of these areas and more. The first immediate difference is that the ionization field emanating from the sources is inhomogeneous. This reveals itself in both the visuals of Figure \ref{fig:same_fgamma} and the extended formation histograms in Figure \ref{fig:formation_hist}. Another key feature of Figure \ref{fig:formation_hist} is that unlike the formation history in G10 and \cite{Bekki08}, MPGC formation is extended in time rather than abruptly ending at one specific redshift. This extended formation can only come about due to the inhomogeneous way in which the ionization field propagates through the bulk IGM. Our models are consistent with the overall picture of extended reionization. For the photon-poorest model (M$1\_512 \_\mathrm{ph}150$), the last source is activated as late as z = 4 whereas in the most photon-rich model (M$1\_512 \_\mathrm{ph}5000$), MPGCs cease forming by z = 7 (see Table \ref{tbl:form_ages}).

\subsection{Globular Cluster Dark Halo Formation Hypothesis}
Whether or not MPGCs originally formed within dark matter halos is still an open question. In this work we do not necessitate that they do but rather ask what sort of estimations can one derive if the dark matter halo formation channel produces distributions similar to observations. We conclude that although some GCs may have formed in dark matter halos and were subsequently stripped upon merging with the central host galaxy, a portion of the Galactic globular cluster population we observe today may have come about by other means (e.g. mergers, giant molecular clouds, tidal tails). In Model 1, even before including effects like dynamical disruption, the radial distributions are either too shallow or don't contain enough GCs to reproduce the Milky Way distribution (even if one assumes several GCs formed per dark matter halo). Though the delayed model (Model 2) does increase the number of potential sources, a great majority of them don't populate the required locations in order to reconcile the simulated distribution with the Milky Way MPGC distribution. This however ignores the contribution from external sources inundating the IGM with ionizing photons which would suppress low redshift GCs. If we adopt Model 2 \textit{and} truncate sources at z  = 10, the spatial distributions are consistent with that of the Milky Way MPGCs. 

\subsection{Convergence}
\label{subsec:convergence}
\begin{figure}
\centering
\begin{tabular}{c}
\includegraphics[width = 0.45\textwidth]{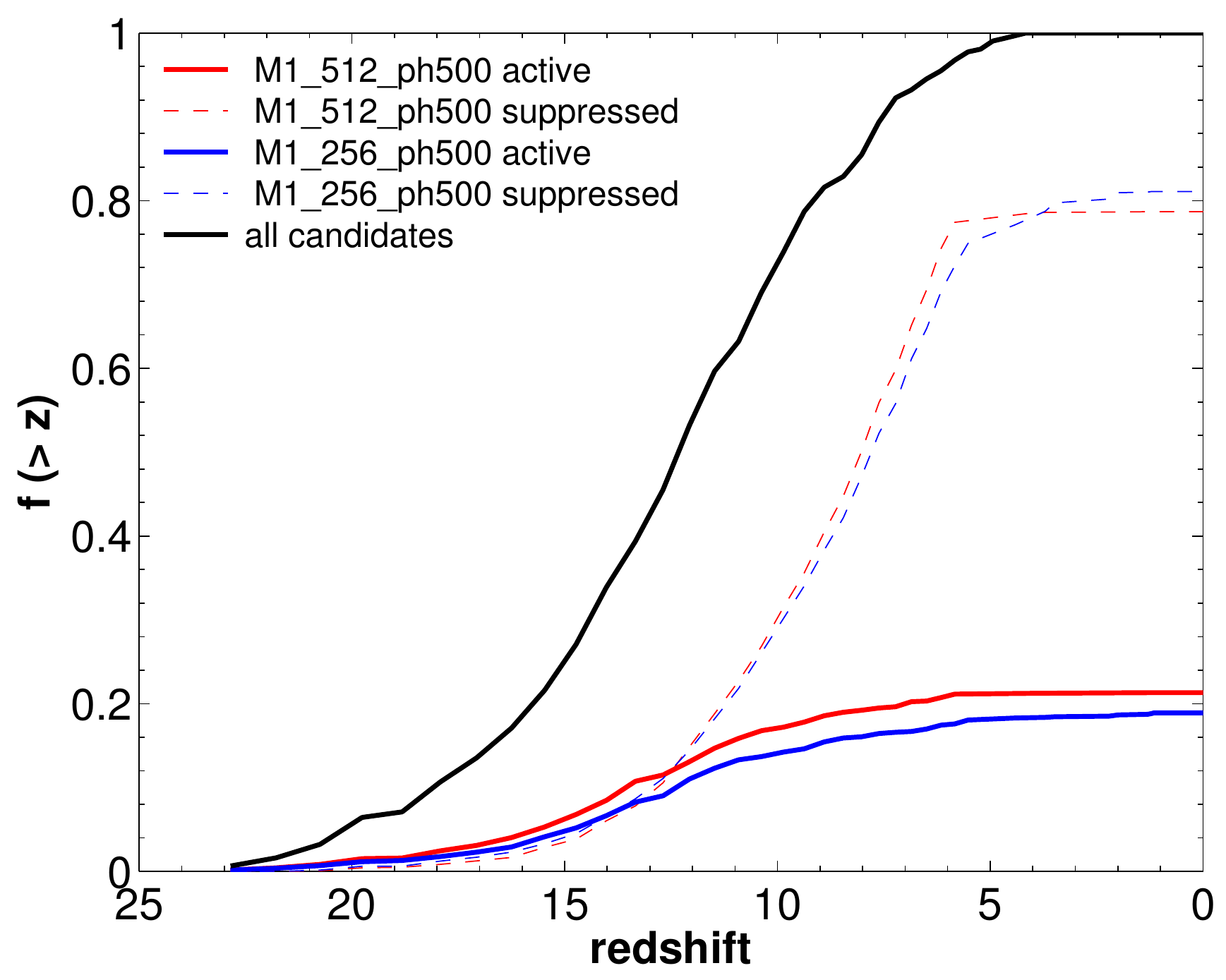}
\end{tabular}
\caption{Convergence test for our fiducial model, M$1\_512 \_\mathrm{ph}500$. In terms of the cumulative formation number, the lower resolution, 256$^3$ mesh shows a near-complete agreement with the higher resolution 512$^3$ mesh. A KS test reveals that the  hypothesis that they are drawn from the same distribution cannot be rejected at the 5$\%$ significance level.}
\label{fig:convergence}
\end{figure}

An important concern for the models is that the results are not dependent on the resolution of the particular simulation used. G10 has already shown that the identification of sources is not resolution-limited by comparing the formation history of objects found within the higher resolution AqA2 halo to the lower resolution AqA3 halo. Here, we are interested in the dependence of the results on the resolution used by {\sc{C$^2$-Ray}}. In conjunction with each of the 512$^3$ suite of models, a lower-resolution 256$^3$ simulation was carried out on our fiducial model (M$1\_512 \_\mathrm{ph}500$) to test for convergence. In Figure \ref{fig:convergence}, we present the formation history of metal-poor GCs in our fiducial model at both resolutions, 256$^3$ and 512$^3$. Both models are in excellent agreement with one another varying only by few percent in number. A KS-test also verifies that they are consistent with having been drawn from the same distribution at the 5$\%$ significance level. This agreement shows that the results are not significantly biased by the resolution of the simulations used.

\subsection{Caveats Of This Work}
\label{subsec:caveats}
Although we have tried to eliminate as many known weaknesses of the models used in this study, there are still a number of remaining caveats which need to be addressed in future studies to improve upon this work.
\begin{itemize}
\item \textit{Accountability of the ionization sources;} whilst the majority of the local ionizing flux will come from low mass dark matter halos in the early Universe as time evolves, each of our models diverges from an accurate account of the \textit{true} ionizing sources. This is because field stars, non-dark matter dominated giant molecular clouds and other such objects will begin to play a dominating role at lower redshifts which are not yet modelled in the current work. Indeed recent work by \cite{Iliev11, Alvarez09} suggests a reionization overlap scenario for the local Milky Way volume whereby reionization was kickstarted by low mass sources locally, but their effective flux is superseded by with the ionizing flux of the Local Group. Nearby proto-clusters (e.g. Virgo and Fornax) clusters would also have had an appreciable affect at later times. When and where this overlap occurred is critical in understanding the origin of present day satellite structures.


\item \textit{Unmerged GCs}; Throughout this work we have limited our sample to include only those which have merged with the central halo by z = 0. This was done to not only ensure an accurate sample but for simplicity. Candidates which do not merge with the central halo by z = 0 could be Galactic dwarf galaxies as they will have retained their dark matter halos, though this is highly speculative. Figure 10 of G10 shows the spatial distribution of these non-merged objects is consistent with those GCs found surrounding neighbouring dwarf galaxies (see \citealt{Mateo98}). Overall there are 47 distinct halos containing 68 GCs at z = 0, indicating that there has been not only major merging with the central host, but also minor-merging of GCs with proto-dwarf galaxies. Whilst there is sufficient research to suggest GCs could have formed in dark matter halos and were subsequently stripped of their outer envelope, it remains unclear where the degeneracy breaks between GC $\&$ dwarf galaxy formation processes. An examination of these non-merged objects needs to be carried out in a more detailed manner to determine their connection (if any) to the known dark matter dominated satellites of Milky Way.

\item \textit{GC formation efficiency}; All of the work presented here assumes that for every dark matter halo, exactly one GC will form in the centre of its potential well. Given that fragmentation of large gas clouds can lead to multiple stellar populations within a single dark matter halo, it still remains unclear as to exactly how many GCs will form in a given dark halo. In any case, it would only shift the halocentric radii plots upwards and not really have any impact on the structural properties of the distributions. Understanding GC formation efficiencies better will lead to better estimates of their overall numbers if they do form via the dark halo formation channel.

\item \textit{Statistical limitation using one Aquarius halo}; Perhaps one of the biggest drawbacks of this work is that it only examines one Milky Way type galaxy. In order to get a better understanding of the underlying physics involved, a larger sample of Milky Way type galaxies would be required.  The statistical variation in the number of satellites in the Aquarius simulation has also been shown to vary by a factor of 2 or more (\citealt{Lunnan12}). The overall conclusion that GCs contributed a non-trivial amount to the reionization of the local IGM would be unlikely to change given that the most conservative of our models reionizes more than 50$\%$ both by mass and volume of the local IGM by z = 10 (see Fig. \ref{fig:vol_ion}). Although the Aquarius suite consists of six such simulations, the logistics behind analysing the remaining five halos was not feasible at this time. 

\item \textit{Treatment of dynamical disruption}; the exact nature of the dynamical processes inside the central host halo after a halo merges is inherently difficult to treat properly. In our model, we trace the single most bound particle of the in-falling MPGCs. These trace particles could be susceptible to violent dynamical changes within the halo resulting in spatial distributions that might not necessarily reflect where the true z = 0 position of the baryonic content ended up. In an improvement to this, one could trace the 10$\%$ most bound particles to get a better hold on the z = 0 position of the clusters under study. Whilst we observe that the disruption process is a critical component in determine the final z = 0 numbers (e.g. as many as half of the original GC population to have formed at high redshift is destroyed by dynamical disruption), a lot more work needs to be done to more accurately determine the degree to which GCs are destroyed over the course of their evolution.

\item \textit{Dependence on ionization cell-fraction threshold}; Since our simulations do not (yet) have gas and do not calculate the temperature, we use the local ionized fraction as a proxy for the temperature (which has a few uncertainties).  At very high redshifts things are reasonably simple since it is typically a simple transition from largely neutral to almost completely ionized as the ionization front overruns the cell in question. A noted problem is what to do after this point. Since the ionization suppresses the low-mass sources, some regions start to recombine (and cool), which eventually should allow the formation of low-mass sources again, but the question is when does this occur? This is a source of uncertainty and so a number of cell fraction thresholds need to be tested in future work. At the moment, the cell fraction ($x_{thresh}$) was set to 0.1 for this work but arguments can be made to raise this threshold to higher values of $x_{thresh}$ = 0.5 or even 0.9 (see Appendix B of Iliev et al. 2011). Model 1 should be retested using these thresholds to determine how much stochasticity one would expect from varying this threshold. 

\item \textit{Clumping}; volume averaged recombination rates in an inhomogeneous IGM scale with the clumping
factor $C = \left<\rho^2\right>/\left<\rho\right>^2$, where the $\rho$ denotes volume averaged densities. In our work, we set C = 1 (no clumping). The dependence of halo suppression for varying degrees of clumping is as yet unknown but will be implemented in subsequent work.
\end{itemize}

Recently, \cite{Spitler12} combined observations  with novel modelling to detect evidence of inhomogeneous reionization by MPGCs on cluster scales. They exploited a fundamental characteristic of galaxy assembly (i.e. spatial biasing and kinematics of MPGCs) to constraint the local reionization epoch around individual galaxies. They found a joint constraint from three galaxies of $z_{reion} = 10.5^{+1.0}_{-0.9}$ which agrees well with the latest WMAP constraint on $z_{reion}$. Interestingly they also found a 1.7$\sigma$ indication that low-density environments were reionized before medium and high-density environments. These results are consistent with the theory that reionization was prolonged in duration with neutral-gas surviving in high-density environments for extended periods. The redshift range of reionization of the local IGM in this work is consistent with their results to within 1$\sigma$.  

Only recently, have computational developments made possible the ability to simulate vast cosmological volumes with exquisite resolution. Future studies such as the work in this paper could be applied to high resolution cluster environments to verify the claims made by \cite{Spitler12}. Preliminary work will be carried out on the Millennium-II simulation but more tailored simulations for this type of study would be desirable.

\section{Conclusions}
\label{sec:conclusion}
Over the course of this work, we have drawn a number of important conclusions regarding the reionization of the inter-galactic medium by metal-poor globular clusters. These conclusions are as follows:
\begin{enumerate}
\item Treating the ionization field in an inhomogeneous manner results in substantially different numbers and spatial distributions. Our basic model adopting this type of treatment produces comparable numbers of GC candidates to that of the Milky Way but results in radial distributions which are too shallow. Whilst this model does not rule out possibility of GCs forming inside dark matter halos at high redshift, it does indicate that GCs must have formed via other mechanisms to account for the lack of numbers at low radii in our model. 

In our so-called `delayed model' where GCs are able to become active after having been previously suppressed, the number of candidate GCs drastically increases. These additional candidates however do not end up residing within the inner, R $<$ 60 kpc, of the host halo at z = 0 and so the resulting radial distributions are still not comparable to that of the Milky Way. 

In the arbitrary truncation model, the radial profiles and numbers are comparable to the Milky Way metal-poor GCs. This is because in the truncation model (a) there are extra sources available at high redshift which boosts the number of sources at small radii and (b) it by definition, totally removes the sources below $z_{\mathrm{trunc}}$ which reduces the number of sources at large radii. This ultimately shows that realistic populations can be produced if external sources dominate the photon budget at later times. 

Overall, the more sophisticated work in this paper clearly shows that treating the ionization of the local IGM in a spatially inhomogeneous manner leads to the IGM being reionized at different places at different times which greatly impacts the radial distributions of MPGCs.

\item Globular clusters injected high numbers of ionizing photons into the IGM at high redshift. Conservative efficiency estimates of $f_\gamma = 150$ photons/baryon would have resulted in MPGCs ionizing more than 50$\%$ of the local IGM mass and 60$\%$ of the local IGM volume by z = 10 (within an enclosed volume of 2$^3$ $h^{-3}$ Mpc$^3$ centred on the host) and upwards of 98$\%$ of the volume and 90$\%$ of the mass for the photon-richest model ($f_\gamma = 5000$ photons/baryon). We also estimate that as many as 10$^{70}$ photons (minimum: 10$^{68.5}$ photons) were injected in the IGM from MPGCs alone during the early build up the Milky Way. Such a quantity of photons would surely impact the formation of structures around not only the Milky Way, but other GC rich environments as well.

\item The suppression rate of MPGCs in our simplest reionization model (M$1\_ 512 \_ \mathrm{ph}150$) was 36 $\%$ of a total 310 possible MPGC sources. The number of objects which form between the photon-poor ($f_\gamma = 150$) and photon-rich ($f_\gamma = 5000$) environments differs by a factor of 2.7.

\item The unsuppressed MPGCs in all models have a narrow age range (mean = 13.34 Gyr, $\sigma$ = 0.04 Gyr) consistent with current ages estimates of the Milky Way MPGCs.

\item The radial distributions for the M$1\_ 512 \_ \mathrm{ph}150$ and M$1\_ 512 \_ \mathrm{ph}250$ models were the only distributions to be statistically consistent with the distribution of the Milky Way MPGCs at the 1$\%$ significance level.

\item In an extension to Model 1, we found that dynamical destruction destroys nearly half (52$\%$) of the total GCs available resulting in a sample of GC candidates which is much smaller in number when compared to the Milky Way metal-poor GCs. A KS-test carried out on all six Model 2 distributions rejected the null hypothesis that they are drawn from the same distribution as the Milky Way MPGCs at  the 1$\%$ significance level.

\item Allowing suppressed halos to turn on once they satisfy the required environmental conditions results in double and some times triple the number of potential GCs having formed. These `delayed GCs' begin forming at z $\sim$ 19 and continue to form up until the present day. The majority of these objects however form within neutral-gas regions of the simulation volume at z $<$ 8 and at R $<$ 300 $h^{-1}$  kpc from the host. Due to their late formation times, these objects are most likely self-enriched and are therefore not suitable for comparison to the Milky Way MPGCs. If we impose truncation at z = 10, the photon-rich radial distributions ($f_\gamma = 1000-5000$ photons/baryon) are consistent with that of the Milky Way.

In summary, a number of avenues could be explored in future work. Measurements of the variation in our results across a wide variety of Milky Way-type halos and a number of other sources of reionization should be studied. Furthermore, testing a range of formation efficiencies, cell fraction thresholds ($x_\mathrm{thresh}$) and an improved treatment of the dynamical disruption will help comprehend the ultimate origin of GCs and the role they played during the epoch of reionization.


\end{enumerate}

\section*{Acknowledgements}

We wish to thank Holger Baumgardt for many useful discussions relating to the dynamical models used in this work.

The simulations for the Aquarius Project were carried out at the Leibniz Computing Center, Garching, Germany, at the Computing Centre of the Max-Planck-Society in Garching, at the Institute for Computational Cosmology in Durham, and on the `STELLA' supercomputer of the LOFAR experiment at the University of Groningen.

This work was supported by funding from the Australian Research Council (grant number DP110102608) and the Science and Technology Facilities Council (grant number ST/F002858/1). ITL and PAT were partially supported by the Science $\&$ Technology Facilities Council (grant number ST/I000976/1). Brendan Griffen would like to acknowledge the support provided by the University of Queensland via a University of Queensland Postgraduate Scholarship.

\label{lastpage}
\bibliography{biblio}
\end{document}